\documentclass[%
 reprint,
showpacs,
 amsmath,amssymb,
 aps,
 pra,
longbibliography,
floatfix,
superscriptaddress
]{revtex4-1}

\usepackage[utf8]{inputenc}
\usepackage[T1]{fontenc}
\usepackage[]{graphicx}
\usepackage{color}
\usepackage[english]{babel}

\usepackage[bookmarks,bookmarksopen,bookmarksdepth=2]{hyperref}
\hypersetup{
    colorlinks=true,
    citecolor=blue,
    linkcolor=blue,
    filecolor=magenta,
    urlcolor=cyan,
}

\usepackage{url}

\usepackage{microtype}
\usepackage{braket}
\usepackage{esvect}
\usepackage[export]{adjustbox}

\usepackage{amsmath,amsfonts,amssymb}
\usepackage{mathtools}

\usepackage[position=top,caption=false]{subfig}

\let\originaleqref\eqref
\renewcommand{\eqref}{Eq.~\originaleqref}

\graphicspath{
{figures/}
}

\begin{document}

\title{Donor spins in silicon for quantum technologies}

\author{Andrea Morello$^{\ast}$}
\author{Jarryd J. Pla}
\affiliation{School of Electrical Engineering and Telecommunications, UNSW Sydney, Sydney, New South Wales 2052, Australia}
\author{Patrice Bertet}
\affiliation{Quantronics Group, SPEC, CEA, CNRS, Universit\'{e} Paris-Saclay, CEA-Saclay, 91191 Gif-sur-Yvette, France}
\author{David N. Jamieson}
\affiliation{School of Physics, University of Melbourne, Melbourne, Victoria 3010, Australia}

\keywords{Spin qubits, Donors in silicon, Ion implantation}

\begin{abstract}
Dopant atoms are ubiquitous in semiconductor technologies, providing the tailored electronic properties that underpin the modern digital information era. Harnessing the quantum nature of these atomic-scale objects represents a new and exciting technological revolution. In this article we describe the use of ion-implanted donor spins in silicon for quantum technologies. We review how to fabricate and operate single-atom spin qubits in silicon, obtaining some of the most coherent solid-state qubits, and we discuss pathways to scale up these qubits to build large quantum processors. Heavier group-V donors with large nuclear spins display electric quadrupole couplings that enable nuclear electric resonance, quantum chaos and strain sensing. Donor ensembles can be coupled to microwave cavities to develop hybrid quantum Turing machines. Counted, deterministic implantation of single donors, combined with novel methods for precision placement, will allow the integration of individual donors spins with industry-standard silicon fabrication processes, making implanted donors a prime physical platform for the second quantum revolution.     

\end{abstract}

\maketitle

\section{\label{sec:introduction}Introduction}

All semiconductor electronic devices use dopants to tailor the electrical properties of the host semiconductor material. For classical applications, doping is an emergent, collective phenomenon that depends on the nature and average density of the dopants. However, the steady miniaturization of electronic devices has naturally brought the behavior of \emph{individual} dopants into the technological spotlight. On the one hand, random fluctuations in the precise number \cite{asenov2003} and location \cite{pierre2010} of dopants can disrupt the operation of nanoscale transistors; on the other hand, gaining \emph{control} over the individual dopants can open the path to improved functionality of classical devices \cite{shinada2005} or radically new applications in quantum science and technology \cite{morton2011}.

At the extreme lower end of the size scale, single dopants were among the first plausible physical systems suggested for use in quantum computing, ever since Kane proposed to encode a quantum bit (qubit) of information in the nuclear spin of a $^{31}$P donor atom in silicon \cite{kane1998}. The appeal of the Kane proposal was in combining the exceptional quantum coherence properties of donor spins in silicon \cite{gordon1958,feher1959,tyryshkin2012,steger2012,saeedi2013} with the prospect of exploiting the semiconductor industry's technological roadmap to miniaturizing silicon devices, dictated by Moore's law \cite{mollick2006}. This has motivated a worldwide effort to fabricate, control and scale up single-spin quantum devices in silicon \cite{zwanenburg2013}. The coherent control of the electron \cite{pla2012} and the nucleus \cite{pla2013} of a single $^{31}$P donor were among the first demonstrations of the feasibility of spin-based quantum information processing in silicon.

Moving to larger dimensions, other proposals have sought the use of small ensembles of spins to encode quantum information in collective spin waves excitations \cite{wesenberg2009}. The spin ensembles are addressed using the microwave magnetic fields produced in superconducting resonators \cite{kubo2010}. These proposals seek to exploit the extraordinary success of circuit-quantum electrodynamics (cQED) in harnessing the quantum nature of solid-state devices coupled to microwave photons \cite{schoelkopf2008}, and extend its scope to the control of spins in the solid state. Current efforts are thus focussing on extending cQED methods to small donors ensembles: initial experiments have already achieved significant goals, such as the long-sought observation of cavity-enhanced relaxation (Purcell effect) in a solid-state spin ensemble \cite{bienfait2016}.

Taken together, single-atom and small-ensemble donor devices provide a broad palette of fundamental properties and potential applications of relevance to the ``second quantum revolution'' \cite{dowling2003}, where the rules of quantum mechanics are explicitly used to develop new technologies. This article discusses the status, challenges and opportunities in using donor spins in silicon for quantum technologies. We provide several examples centered around the past work and future plans of the present authors, and integrate them with a discussion of broader proposals for scale-up technologies that are particularly suited to the fabrication pathway based upon single-ion implantation. An overview of the scanning-tunnelling microscope (STM) lithography fabrication pathway is provided in Ref.~\cite{ruess2007}, and recent results on the operation and integration of such donor spins systems in Refs.~\cite{he2019,koch2019spin}. A general introduction to silicon quantum electronics can be found in Ref.~\cite{zwanenburg2013}, and a review of silicon qubits is given in Ref.~\cite{ladd2018}. A broad overview of semiconductor qubits in given in Ref.~\cite{chatterjee2020semiconductor}.

\section{\label{sec:fab}Fabrication and operation of donor-based quantum devices}

\subsection{Single-donor spin qubit devices}  \label{sec:single_donors}

The compatibility with classical semiconductor manufacturing methods is a very appealing aspect of single donors in silicon as platforms for novel quantum technologies. In particular, since ion implantation is the industrial method of choice for doping semiconductors \cite{lecuyer2009}, it is desirable to design quantum devices where ion-implanted donors are integrated with metal-oxide-semiconductor (MOS) nanostructures, i.e. precisely the way in which all modern integrated semiconductor chips are mass-produced.

A prototypical quantum electronic device inspired by classical transistors is the gate-defined silicon MOS quantum dot \cite{angus2007}, whose layout and fabrication is that of a textbook-example MOSFET, augmented with two barrier gates to isolate a many-electron quantum dot. This device exhibits a highly nonlinear current-voltage characteristic with the typical Coulomb conductance peaks, and can thus serve as a single-electron transistor (SET) charge sensor \cite{devoret2000}. 

\begin{figure}[tb]
  \includegraphics[width=8.2cm]{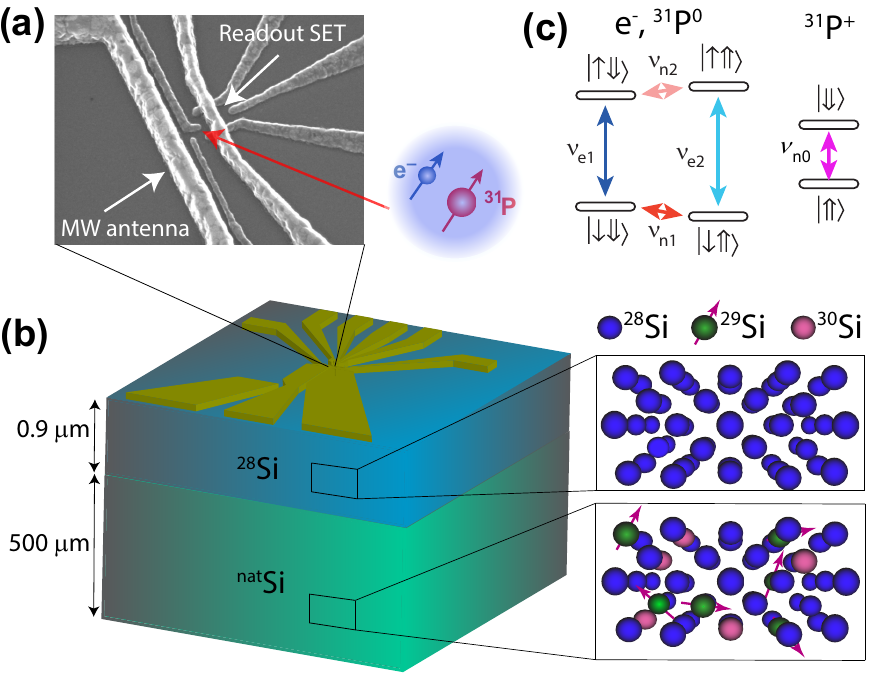}
  \caption{Device layout for single-donor spin qubits. (a) Scanning electron micrograph of a typical single-donor spin qubit device, comprising aluminum gates with 30~nm typical width, fabricated on top of a thin ($5-10$~nm) SiO$_2$ dielectric, grown on top of a silicon wafer. (b) In all our experiments since 2014, we have fabricated the devices on an isotopically-enriched $^{28}$Si epilayer (with residual 800 ppm $^{29}$Si concentration), deposited on a natural silicon wafer. (c) Schematic energy level diagram of a neutral $^{31}$P donor comprising an electron (e$^-$) and a neutral nucleus ($^{31}$P$^0$), or an ionized donor comprising solely the positively charged nucleus ($^{31}$P$^+$). Arrows indicate the electron spin resonance transitions, at frequencies $\nu_{\rm e1,2} = \gamma_{\rm e} B_0 \mp A/2$, and the nuclear magnetic resonance frequencies $\nu_{\rm n1,2}=A/2 \pm \gamma_{\rm n} B_0$ or $\nu_{\rm n0} = \gamma_{\rm n} B_0$ in the ionized case. Adapted from \cite{muhonen2014}.
  \label{fig:donor_device}
  }
\end{figure}

An integrated device architecture for control and readout of individual donors \cite{morello2009} comprises an SET fabricated in close proximity (at a typical distance $\sim 20$~nm \cite{mohiyaddin2013}) to a region containing the implanted donors. The structure is completed by an on-chip broadband microwave antenna \cite{dehollain2012}, optimized to deliver oscillating magnetic fields at the donor location, in order to perform nuclear and electron spin resonance (Figure~\ref{fig:donor_device}). Significantly, the integration of gate-defined naoelectronic devices with ion-implanted donors has been demonstrated within a complete, foundry-based CMOS process flow \cite{singh2016}. It was also shown that the donor implantation does not impact the charge stability of the devices \cite{rudolph2019}.

The response of donors spins to external fields is described by the Hamiltonian (in frequency units):

\begin{equation} \label{eq: H_donor}
\begin{split}
\mathcal{H} = \left(\gamma_\mathrm{e}S_z-\gamma_\mathrm{n}I_z\right)B_0+A \hspace{0.25em} \mathbf{S}\cdot \mathbf{I} + \mathcal{H}_\mathrm{Q}+\\
 + \left(\gamma_\mathrm{e}S_x-\gamma_\mathrm{n}I_x\right)B_1 \cos \left( 2 \pi f t\right) , 
\end{split}
\end{equation}

where $\gamma_e \approx 28$~GHz/T and $\gamma_n$ are the electron and nuclear gyromagnetic ratios, $\mathbf{S}=(S_x \hspace{0.5em} S_y \hspace{0.5em} S_z)^{\rm T}$ and $\mathrm{I}=(I_x \hspace{0.5em} I_y \hspace{0.5em} I_z)^{\rm T}$ are the electron and nuclear vector spin operators, and $A$ is the electron-nuclear hyperfine coupling strength. The relevant parameter values for group-V donors in silicon are given in Table~\ref{tab:donor_parameters}. $B_0$  is a static external magnetic field (typical values are in the $\sim 0.2 - 1.5$~T range, corresponding to electron spin resonance frequencies $\nu_{\rm e} \sim 6 - 40$~GHz) that produces a Zeeman energy splitting, and $B_1$  is an oscillating magnetic field (of typical strength $\sim 1-100$~$\mu$T) applied to induce coherent transitions between the eigenstates of the static part of the Hamiltonian. If, as is typically the case, the electron Zeeman interaction $E_{\rm Z} = \gamma_{\rm e} S_z B_0$ dominates over all the other terms, the Hamiltonian eigenstates are to a good approximation the tensor products of the basis states of the electron spin, $\ket{\uparrow},\ket{\downarrow}$, and those of the nuclear nuclear spin, $\{\ket{m_I}\}, m_I = -I, -I+1, \ldots I$. $\mathcal{H}_{\rm Q}$ is the nuclear quadrupole interaction, which can exist in nuclei with spin $I>1/2$, and is discussed in more detail in Section~\ref{sec:quadrupole}.

\begin{table}

\caption{\label{tab:donor_parameters} Hamiltonian parameters for group-V donors in silicon \cite{mourik2018}.}
\begin{tabular}{ccccc}
\hline
Donor & $I$ & $A$ & $\gamma_\mathrm{n}$ & $Q_\mathrm{n}$ \\
& & (MHz) & MHz/T & $(10^{-28} \mathrm{m}^2)$ \\
\hline
$^{31}$P & $1/2$ & 117.53 & 17.26 & - \\
$^{75}$As & $3/2$ & 198.35 & 7.31 & 0.314\\
$^{121}$Sb & $5/2$ & 186.80 & 10.26 & [-0.36 , -0.54]\\
$^{123}$Sb & $7/2$ & 101.52 & 5.55 & [-0.49 , -0.69]\\
$^{209}$Bi & $9/2$ & 1475.4 & 6.96 & [-0.37 , -0.77]\\
\hline
\end{tabular}

\end{table}

Coherent control of the donor spin states is usually obtained by magnetic resonance \cite{pla2012,pla2013}, via the term $\left(\gamma_{e}S_x-\gamma_\mathrm{n}I_x\right)B_1 \cos \left( 2 \pi f t\right)$ (see, however, Sections~\ref{sec:multi} and \ref{sec:quadrupole} for cases where the spins are controlled electrically). Performing magnetic resonance on a single spin is not, in general, any harder than doing so on a spin ensemble. A nanofabricated coplanar-waveguide antenna \cite{dehollain2012} easily delivers oscillating fields $B_1 \sim 100$~$\mu$T at 40 GHz, within an order of magnitude of the values produced by conventional ESR spectrometers that use bulk cavities. The true challenge in measuring single or few spins lies in the readout method.

For single donors, the electron spin state can be read out in single-shot and with high fidelity (typically $>90$\%) using spin-to-charge conversion \cite{morello2010}. This method, first pioneered in quantum dots \cite{elzerman2004} and charge traps \cite{xiao2004}, exploits the tunnel coupling between the electron charge of the spin under measurement and a cold charge reservoir at temperature $T_{\rm el}$, when the electron Zeeman energy $E_{\rm Z}$ exceeds the thermal broadening $\approx 5 k_{\rm B}T_{\rm el}$ ($k_{\rm B}$ is the Boltzmann constant). In these conditions, one can electrostatically tune the system in such a way the the electrochemical potential of the spin-up state, $\mu_{\uparrow}$, lies above the Fermi energy $E_{\rm F}$ of the charge reservoir, while the spin-down electrochmical potential $\mu_{\downarrow}$ lies below it. As a consequence of the Pauli exclusion principle, the spin-down electron is forbidden from tunneling out of the donor, since it faces occupied electron states in the charge reservoir. Conversely, a spin-up state is able to tunnel into an empty electron state of the reservoir, leaving behind the positive charge of an ionized donor. The presence of such a charge can be detected, in a time-resolved manner, by the nearby SET that acts as a charge detector \cite{morello2009}. In all successful single-donor experiments conducted so far, the island of the SET also served as the cold charge reservoir for spin-dependent tunneling. Typical SET charge sensitivities easily reach $10^{-5}$~$e/\sqrt{\mathrm{Hz}}$ ($e$ is the electron charge), which translate into the ability to detect the tunneling of a single electron in real-time with a bandwidth $\sim 100$~kHz \cite{morello2010}. A very useful feature of this method is that, at the end of the readout phase, the electron is always reset to the $\ket{\downarrow}$ state, either because it was in that state to being with and never left the donor, or because the $\ket{\uparrow}$ electron tunnelled out and was replaced by a $\ket{\downarrow}$ electron extracted from below the reservoir Fermi level.

The readout of a donor nuclear spin is obtained by using the electron spin as an ancilla \cite{pla2013}. The hyperfine coupling $A \hspace{0.25em} \mathbf{S}\cdot \mathbf{I}$ ensures that the frequency $\nu_{\rm e}$ at which the electron responds to an oscillating magnetic field depends on the state $\ket{m_I}$ of the nucleus: $\nu_{\rm e}(m_I) \approx \gamma_{\rm e}B_0 + A m_I$. The nucleus is thus projectively measured in the state $\ket{m_I}$ if the electron spin, initially set in the $\ket{\downarrow}$ state, is excited to the $\ket{\uparrow}$ state by by a $\pi$-pulse of oscillating field at frequency $\nu_{\rm e}(m_I)$. Moreover, this method of nuclear measurement is almost perfectly quantum nondemolition (QND) \cite{braginsky1996}: since $E_{\rm Z}$ is by far the dominant term in the Hamiltonian, the hyperfine coupling can be approximated with $AS_zI_z$, thus complying with the QND condition that the interaction Hamiltonian commutes with the Hamiltonian of the system under measurement. This allows the use of repetitive QND readout, i.e. the repetition of the cycle (initialize $\ket{\downarrow}$ -- $\pi$-pulse at $\nu_{\rm e}(m_I)$ -- electron readout) $N$ times to improve the nuclear readout fidelity. With $N\sim 100$, the nuclear readout fidelity easily exceeds 99.8\%, limited by electron-nuclear cross-relaxation \cite{pla2013}.

\subsection{Engineered devices for coherent control of donor ensembles}  \label{sec:ensemble_devices}

Ensemble-based donor technologies exploit the excellent coherence properties of donor spins, without the requirement for individual donor addressability and readout. This can be achieved by inductively coupling the ensemble to high-quality-factor superconducting resonators (Section~\ref{sec:ensembles}). These hybrid devices may also benefit from a MOS-compatible fabrication process: ion implantation is first used to introduce donors to the silicon substrate and a high-quality superconducting film is then deposited on the sample via magnetron sputtering or electron-beam evaporation. The precise choice of superconducting film depends on the species of donor employed. Bismuth ($^{209}$Bi) has the largest hyperfine interaction of the group-V impurities and also possesses the greatest nuclear spin (see Table~\ref{tab:donor_parameters}), providing a zero-field splitting of 7.375 GHz – a frequency matching the typical range for superconducting quantum electronics and therefore compatible with the low critical magnetic field of aluminum \cite{bienfait2015}. For other donors ($^{31}$P, $^{75}$As, $^{121/123}$Sb), magnetic fields of several hundred millitesla are required to produce Zeeman energies of order 10 GHz and superconducting materials with larger critical fields are selected (Nb, NbN, TiN, NbTiN) \cite{sigillito2014}.

The simplest Hamiltonian describing the coupling of an ensemble of $N$ donor spins to a resonator is $H_I=\sqrt{N}g_0(aS_++a^\dag S_-)$ \cite{wesenberg2009, kubo2010}, where $g_0$ is the coupling strength of a single donor to the superconducting resonator, $a$ ($a^\dag$) is the annihilation (creation) operator of the resonator and $S_+$ ($S_-$) describes the creation (annihilation) of an excitation in a collective spin mode called a spin wave. The spin-wave-photon coupling is therefore enhanced by a factor $\sqrt{N}$ compared to the single donor interaction strength $g_0$. Coherent transfer of quantum information between the systems is achieved by engineering devices in the high cooperativity limit $C=g_{ens}^2/\kappa\Gamma\geq\ 1$, where the effective ensemble coupling $g_{ens}=\sqrt{N}g_0$ overcomes the effects of system loss, namely the resonator photon decay rate $\kappa$ and the dephasing rate $\Gamma$ of the spin-wave, caused by the inhomogeneous broadening of the spin ensemble \cite{kubo2011}.

\begin{figure}[tbph]
  \includegraphics[width=8.2cm]{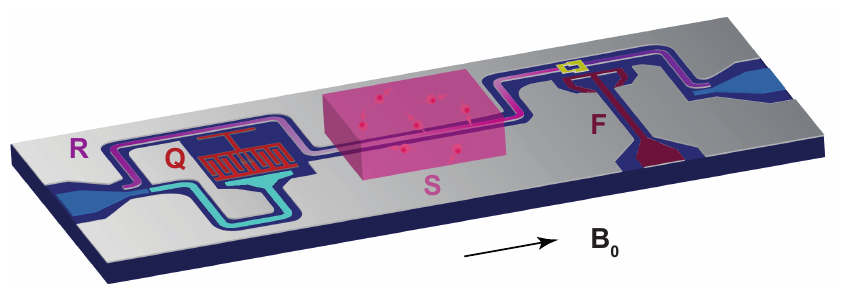}
  \caption{\label{fig:hybriddevice} A typical hybrid device consisting of a superconducting resonator (R) coupled to an ensemble of donors in silicon (S) and a transmon qubit (Q). The resonator has an embedded superconducting quantum interference device (SQUID) and flux line (F) to allow for frequency tunability. This particular device implements a quantum memory (Section~\ref{sec:ensembles}), where quantum states from Q are transferred to S via R. Figure adapted from \cite{kubo2011}.
  }
\end{figure}

The donor ensemble is addressed by performing microwave measurements of the superconducting resonator. A microwave signal, in the form of a weak coherent state or single-photon quantum state (e.g. originating from a quantum bit), is sent to the resonator and may be efficiently absorbed by a coupled donor ensemble under certain conditions \cite{ afzelius2013, julsgaard2013}. Once retrieved from the ensemble, the weak microwave signal is sent through a series of amplification stages at different temperatures (typically $\sim$~10~mK, 4~K and $\sim$~300~K) in order to maximise the signal-to-noise ratio of the measurement, before undergoing homodyne demodulation and detection at room temperature. A typical hybrid device architecture is depicted in Figure~\ref{fig:hybriddevice}.

\section{\label{sec:benchmarks}Performance benchmarks}

Donor spins are among the most coherent quantum systems in the solid state. This results from a combination of weak spin-orbit coupling in the silicon conduction band \cite{zwanenburg2013}, and the large natural abundance of spin-zero nuclear isotopes of silicon \cite{itoh2014}, with only 4.7\% $^{29}$Si carrying nuclear spin $1/2$. It was known since the late 1950s that the isotopic enrichment of the spinless $^{28}$Si isotope would further improve spin coherence \cite{gordon1958}. In more recent times, thanks to the availability of ultra-enriched $^{28}$Si crystals with $<50$~ppm $^{29}$Si originating from the Avogadro project \cite{becker2010}, ensemble spin resonance measurement have shown exceptional coherence times, exceeding seconds for the donor-bound electron \cite{tyryshkin2012} and hours for the nuclear spin of ionized $^{31}$P donors \cite{saeedi2013}.

These results have been, to some extent, replicated in single-donor devices, although the use of different cryomagnetic systems, the presence of metallic electrodes near the donor, and the use of less aggressively enriched $^{28}$Si (with typically 800~ppm residual $^{29}$Si) have so far prevented reaching the extreme values found in bulk.

Table \ref{tab:benchmarks} summarizes the key performance metrics of single-donor spins: the spin-lattice relaxation time $T_1$, the pure depahsing time $T_2^*$ as obtained from Ramsey experiments, the Hahn echo coherence time $T_2^{\rm H}$, the coherence time using Carr-Purcell-Meiboom-Gill (CPMG) dynamical decoupling $T_2^{\rm CPMG}$, the time to perform a $\pi$ rotation ($X$ gate), the single-shot readout fidelity $\mathcal{F}_{\rm meas}$ and the fidelity of single-qubit Clifford gates $\mathcal{F}_{\rm Clifford}$. The coherence time $T_2^{\rm CPMG} = 35.6$~s of the ionized $^{31}$P$^+$ nucleus \cite{muhonen2014} was until recently \cite{bradley2019} the record value for any single-qubit in the solid state. Also notable is the demonstration of electron-nuclear entanglement verified via the violation of the Clauser-Horne-Shimony-Holt version of Bell's inequality, with a maximum Bell signal $S = 2.70(9)$ \cite{dehollain2016bell}, very close to the theoretical maximum $S=2\sqrt{2}$ and representing the highest Bell signal observed in the solid state to date.

The single-qubit Clifford gate fidelities exceed 99.9\% \cite{muhonen2015,dehollain2016} and thus surpass the threshold for fault-tolerance operation in certain error correction schemes, such as the surface code \cite{fowler2012}. As in all other semiconductor-based quantum technologies, the main focus and challenge for the near future is achieving similarly high fidelities for 2-qubit logic gates.

\begin{table}

\caption{\label{tab:benchmarks} Performance metrics for the spin of electron ($e^-$), neutral nucleus ($^{31}$P) and ionized nucleus ($^{31}$P$^+$) of a single phosphorus donor implanted in a silicon metal-oxide-semiconductor device. $^{\dag}$For the nuclear spin, $T_1$ is to be understood as the flipping time caused by repeated donor ionization. The `true' spin-lattice relaxation time is astronomically long.}

\begin{tabular}{llll}
\hline
Metric & $e^-$ & $^{31}$P & $^{31}$P$^+$ \\
 \hline
$T_1$ (s) & 9.8(7) \cite{tenberg2019} & $\gg 100^{\dag}$ \cite{pla2013} & - \\
$T_2^*$ (ms) & 0.27 \cite{muhonen2014} & 0.570 \cite{muhonen2014} & 600 \cite{muhonen2014}\\
$T_2^{\rm H}$ (ms) & 1.1 \cite{muhonen2014} & 20 \cite{muhonen2014} & 1800 \cite{muhonen2014} \\
$T_2^{\rm CPMG}$ (s) & 0.55 \cite{muhonen2014} & 0.02 \cite{muhonen2014} & 35.6 \cite{muhonen2014}\\
$\pi$-pulse ($\mu$s) & 0.15 \cite{pla2012} & 25 \cite{pla2013} & 30 \cite{pla2013}\\
$\mathcal{F}_{\rm meas}$ & 92\% \cite{morello2010} & 99.8\% \cite{pla2013} & 99.8\% \cite{pla2013} \\
$\mathcal{F}_{\rm Clifford}$ & 99.94\% \cite{dehollain2016} & -  & 99.98\% \cite{muhonen2015} \\
\hline
\end{tabular}

\end{table}

\section{\label{sec:multi}Multi-qubit operations}

Multi-qubit quantum logic gates require a physical interaction between the qubits. Below we discuss the most promising routes to demonstrating high-fidelity 2-qubit logic gates with donors in silicon.

\subsection{Exchange interaction} \label{sec:exchange}

The two natural forms of physical interaction between donor electron spins are the magnetic dipole coupling \cite{desousa2004} and the exchange interaction \cite{koiller2001}. The weakness of the magnetic dipole coupling is expected to result in slow ($\sim 1$~ms) 2-qubit logic gates \cite{hill2015,ogorman2016}. Conversely, the exchange interaction strength can exceed 1~GHz \cite{dehollain2014,gonzalez2014}, and thus mediate sub-nanosecond exchange oscillations \cite{he2019}. The dynamic control of $J$ yields a $\sqrt{i\mathrm{SWAP}}$ gate \cite{petta2005}, which is the native 2-qubit entangling logic operation when the qubits' coupling is stronger than their energy detuning.

The original proposal by Kane~\cite{kane1998} advocated the use of a locally-gateable, strong exchange interaction to mediate 2-qubit logic gates. This requires the placement of a narrow, well-aligned ``J-gate'' between  a pair of donors. Even to this day, this remains an exceptionally challenging fabrication requirement, since the typical inter-donor distance would have to be of order 10-15~nm. STM fabrication methods can produce all-epitaxial gates on this size scale and placement precision, but such gates are within the silicon bulk, i.e. not isolated from the donors by a dielectric, which requires placing them $\approx 50$~nm away from the spins. Therefore, even in STM-fabricated devices, dynamical control of the exchange interaction is preferentially achieved by detuning the electrochemical potential of the two spins using side gates~\cite{he2019}.

In the opposite limit, where $J$ is weaker than the detuning, the native 2-qubit operation becomes the Controlled-Rotation (CROT) gate \cite{zajac2018,huang2019}, which is equivalent to the well-known Controlled-NOT (CNOT) gate with an additional $\pi/2$ phase shift. A CROT gate can be obtained by a simple ESR $\pi$-pulse, if the resonance frequency of the target qubit is conditional on the state of the control qubit. This gate does not require dynamic control of the exchange coupling, and therefore no need to fabricate a ``J-gate'' between the donors. For a pair of donor electron spin qubits, a large energy detuning can be introduced naturally and conveniently by preparing the nuclear spins in opposite states \cite{kalra2014}. The detuning then equals the hyperfine coupling $A$. High-fidelity CROT operations can be achieved for values of exchange coupling $J$ in the very broad range comprised between the inhomogeneous resonance linewidth $\Delta \nu = \mathrm{ln}(2)/(\pi T_2^*) \sim 10$~KHz and the hyperfine coupling $A \sim 100$~MHz (Table \ref{tab:donor_parameters}), making this type of 2-qubit logic gate ideal for ion-implanted donors, where the relative position of the donors can only be controlled to within a few nanometres (Table~\ref{tab:ions}). 

\begin{figure}[tbph]
  \includegraphics[width=6cm]{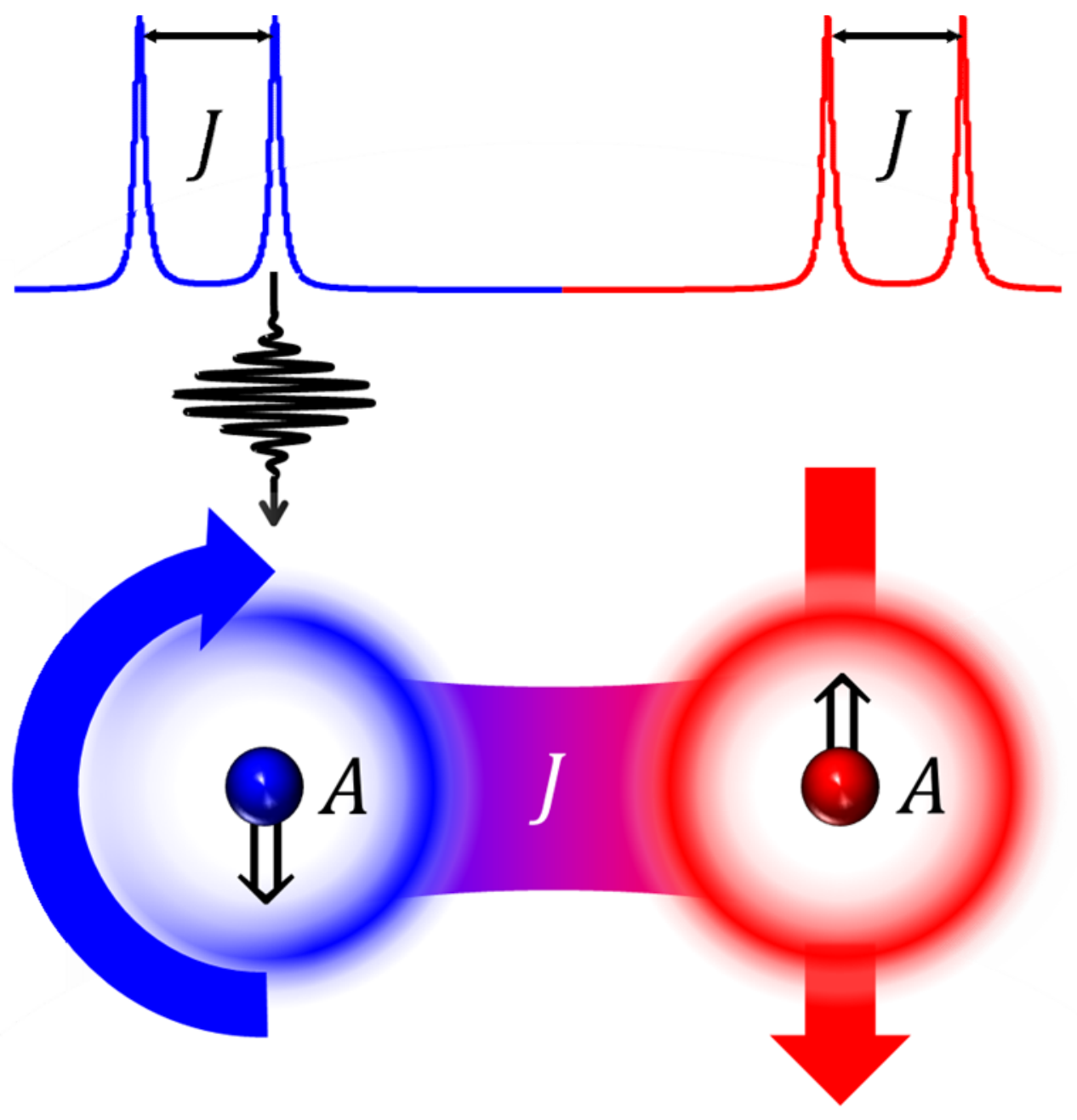} 
  \caption{Sketch of the working principle of a resonant CROT 2-qubit gate between donor electron spins, mediated by the exchange interaction $J$ and controlled by the hyperfine coupling $A$. When $J<A$, the electron spin resonance spectrum consists of six resonance lines, four of which (depicted in the drawing at the top) appear in pairs separated by $J$. A $\pi$-pulse on each of these resonances constitutes the rotation of one qubit conditional on the state of the other, i.e., a 2-qubit CROT gate.
  }
  \label{fig:Jgate}
\end{figure}

In a recent experiment with randomly-implanted donors, we demonstrated the coherent conditional operation of a pair of $^{31}$P donor electron spin qubits in the weak-exchange limit~\cite{madzik2020conditional}. We spectroscopically measured the exchange interaction $J=32.06 \pm 0.06$~MHz, and performed coherent Rabi oscillations on both conditional and unconditional resonance lines. For the conditional resonances, a $\pi$-pulse corresponds to a two-qubit CROT gate~(Fig.~\ref{fig:Jgate}).

\subsection{Flip-flop qubit: electric dipole interaction} \label{sec:flipflop}

Regardless of the choice of native gate, using the exchange interaction always requires the donors to be placed within 15~nm or less from each other, posing a significant challenge to the fabrication tolerances and the size of the classical structures needed for control, readout and interconnects \cite{vandersypen2017}. To circumvent this challenge, we have proposed a new encoding scheme, called the `flip-flop' qubit \cite{tosi2017}, where quantum information is stored in the $\ket{\uparrow\Downarrow},\ket{\downarrow\Uparrow}$ electron-nuclear states of a single donor. Selection rules forbid magnetic dipole transitions between $\ket{\uparrow\Downarrow}$ and $\ket{\downarrow\Uparrow}$. However, the hyperfine interaction $A \hspace{0.25em} \mathbf{S}\cdot \mathbf{I}$ appears as a transverse term in the flip-flop subspace, and therefore flip-flop transitions can be induced by a time-dependent electrical modulation of $A$, at the frequency corresponding to the flip-flop energy splitting $\epsilon_{\rm ff}= \sqrt{(\gamma^+ B_0)^2 + A^2}$, with $\gamma^+ = \gamma_{\rm e} + \gamma_{\rm n}$. Electric field control (Stark shift) of the hyperfine coupling has been demonstrated experimentally in single-donor implanted devices \cite{laucht2015}, and is theoretically well understood \cite{rahman2007}. It requires applying a strong vertical electric field to the donor to displace the electron from the nucleus \cite{calderon2006}, which results in a reduction of $A$. 

The predicted 1-qubit gate time for the flip-flop qubit, $T_{\rm 1q} \approx 50$~ns, depends on the sensitivity of the hyperfine coupling to the oscillating electric field, $\partial A/ \partial E_{\rm ac}$. Such sensitivity can be greatly enhanced by applying a strong dc electric field, displacing the donor-bound electron halfway to the interface with a dielectric. The key feature of the flip-flop qubit is that displacing the electron away from the positively-charged nucleus creates a strong (of order 100 Debye) electric dipole. This dipole is oriented vertically, i.e. perpendicular to the metal gates, and is therefore not screened. Full Hamiltonian calculations reveal that two flip-flop qubits can couple via electric dipole interactions with a strength $\sim 10$~MHz at a distance of 200~nm \cite{tosi2017}. This is conveniently compatible with the typical gate size and pitch of modern CMOS transistor processes \cite{fischer2015} and, since the dipole interaction only decays as the third power of the distance, nanometre-level uncertainties in donor location have negligible effect. 

A two-dimensional array of flip-flop qubits spaced by 200~nm would contain $25\times10^6$ physical qubits per millimeter square: the combination of manufacturability with small size is a key featured of this system, particularly in the context of reaching the number of physical qubits necessary to perform useful, fault-tolerant quantum computation in schemes such as the surface code \cite{fowler2012}.

Despite the presence of a strong electric dipole, the flip-flop qubit is predicted to have long dephasing times, $T_2^* \sim 10$~$\mu$s, thanks to a `second-order clock transition', i.e. a particular region of parameters where the qubit energy splitting is insensitive to the first as well as second derivative of the vertical electric field \cite{tosi2017}. With realistic models for the amplitude and spectrum of the charge noise in silicon devices, we predicted 1-qubit gate fidelities in the 99.9\% range, and 2-qubit $\sqrt{i\mathrm{SWAP}}$ fidelities around 99\%.

The realization of flip-flop qubits and their coupling via electric dipole interactions is highly specific to ion-implanted donor systems, since it requires a hyperfine-coupled nuclear spin, a large polarizability of the electric charge (ruling out other prominent electron-nuclear systems such as nitrogen-vacancy centres in diamond and other spin-carrying color centres \cite{awschalom2018}), and a high-quality dielectric layer $\sim 10$~nm above the donor (ruling out architectures that do not incorporate a gate oxide to isolate the donors from the metallic gates \cite{ruess2007}).

\begin{figure*}[tbph]
  \includegraphics[width=14cm]{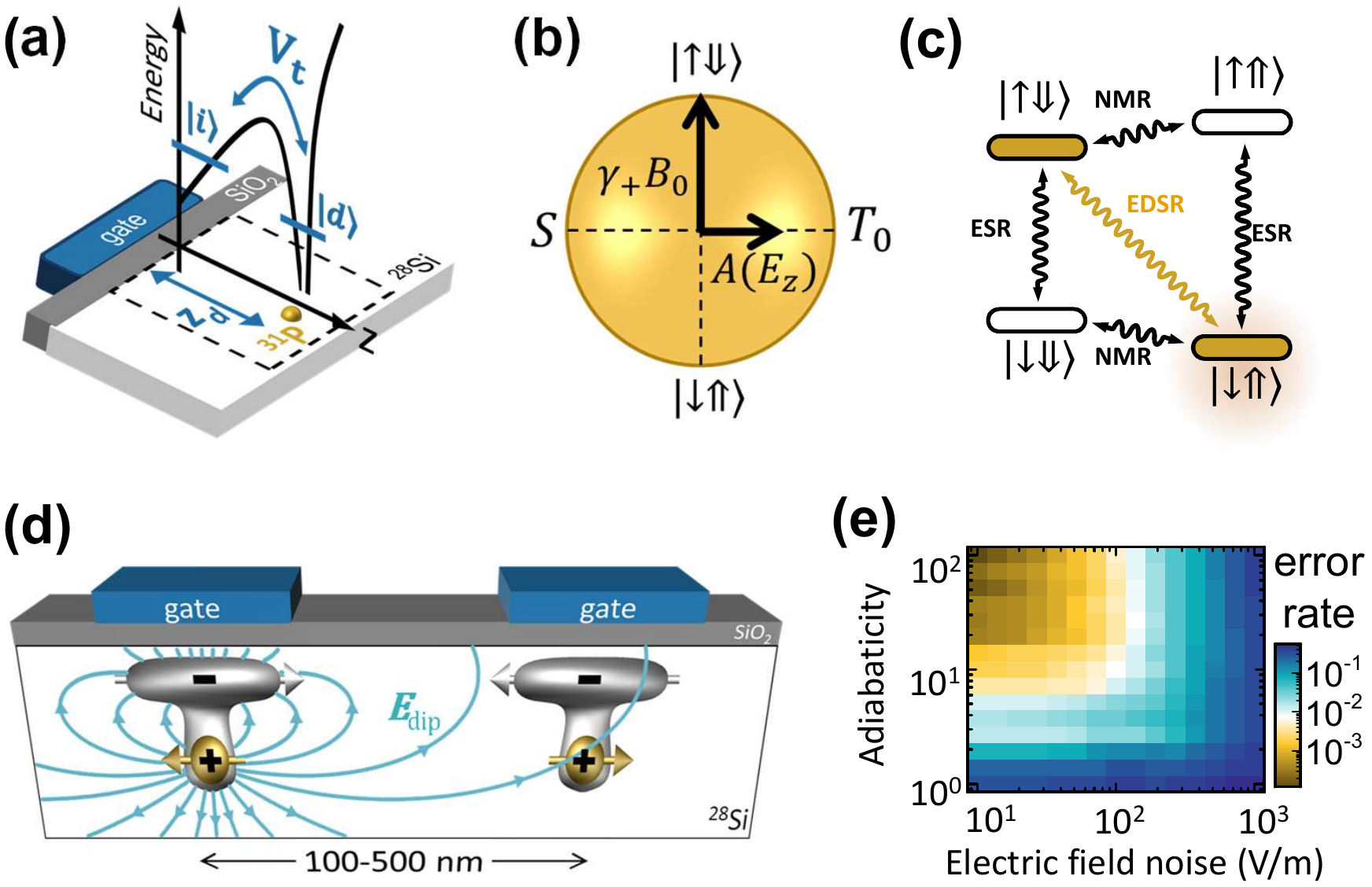} \label{fig:flipflop}
  \caption{Flip-flop qubit and electric dipole coupling. (a) The electron of a donor implanted $\approx 15$~nm under a Si/SiO$_2$ dielectric behaves like a charge qubit, with basis states $\ket{d}$ (electron bound to the donor) and $\ket{i}$ (electron at the interface). (b) The electron-nuclear hyperfine coupling $A$, which depends on the vertical electric field $E_Z$ through the charge qubit state, appears as a transverse term in the Bloch sphere defined by the flip-flop basis states $\ket{\uparrow\Downarrow},\ket{\downarrow\Uparrow}$. (c) The flip-flop qubit is controlled via electric-dipole spin resonance (EDSR) through the electrical modulation of the hyperfine coupling. (d) Displacing the electron charge away from the donor nucleus creates a large electric dipole, that allows coupling two flip-flop qubits at distances $\sim 100 - 500$~nm. (e) The error rate of a 2-qubit $\sqrt{i\mathrm{SWAP}}$ gate is predicted to be below 1\% for electric field noise of r.m.s. amplitude $< 100$ V/m. Adapted from Ref.~\cite{tosi2017}.
  }
\end{figure*}

\section{\label{sec:scaling}Scaling up and spacing out donor qubits}

The issue of spacing out the qubits to allow enough space for interconnects and control electronics is ubiquitous in spin-based devices. We discuss below a few proposals -- not unique to donor qubits, but applicable to them -- aimed at spacing electron spin qubits across distances $\gg 100$~nm.

\subsection{Coupling via spin chains}

An odd-number chain of $N$ spins, coupled by a strong antiferromagnetic exchange interaction $J$, possesses a doubly-degenerate ground state that effectively behaves like a delocalized spin-1/2 system, extending along the length of the chain. The energy gap to the next manifold of excited states is of order $J/N$ \cite{friesen2007} and must be much larger than all other energy scales (spin qubit energy splitting, spectral width of control fields, etc.). Individual qubits, e.g. placed at opposite ends of the chain, can be coupled to the chain by a switchable exchange coupling. Quantum information can be transferred between the qubits at the opposite ends of the chain by two sequential SWAP gates (qubit 1 $\rightarrow$ chain, chain $\rightarrow$ qubit 2), or by a smooth, adiabatic control of the couplings between the chain and the qubits \cite{oh2013}.

A quantitative analysis specific to chains of implanted donors \cite{mohiyaddin2016} reveals that, when the donors are in a bulk-like state, the requirement of large intra-chain $J$ imposes very short ($< 7$~nm) donor distance. However, displacing the electrons from the donors to the quantum dots that naturally form at the Si/SiO$_2$ interface in the presence of the donor image charge \cite{calderon2006}, allows increasing the distance to $\sim 15 - 20$~nm and making the length of the chain exceed 100~nm with just 5 or 7 donors. 

\subsection{Coupling via interface quantum dots}

Ion-implanted donors are naturally placed and operated at a small depth (typically $10-20$~nm) beneath a Si/SiO$_2$ interface. At that interface, one can naturally form single-electron quantum dots that are themselves excellent spin qubits \cite{yang2019silicon,huang2019}. The proximity of donors to interface dots can be exploited in a number of ways.

We have described the flip-flop qubit in Section~\ref{sec:flipflop}, which is a donor-dot hybrid with one electron shared between the two sites. An alternative arrangement is a singlet-triplet qubit, where two (or an even number of) electrons are shared between donor and interface dot \cite{harvey2017coherent}. The operation of singlet-triplet qubits requires both an exchange interaction, and a magnetic field gradient across the two spins \cite{petta2005}. In a donor-dot system, the presence of a nuclear spin with hyperfine coupling $A$ imparts a frequency detuning $A/2$ on the donor-bound electron, equivalent to a magnetic field gradient. With this system it is thus possible to observe coherent singlet-triplet oscillations at frequency $\approx A/2$ \cite{harvey2017coherent}, without resorting to dynamic nuclear polarization of a mesoscopic nuclear spin bath \cite{foletti2009universal}, or to micromagnet structures fabricated on the chip \cite{wu2014two}. Having established a donor-dot singlet-triplet qubit, it becomes possible to envisage coupling multiple such qubits via their electrostatic interaction, as demonstrated earlier in GaAs/AlGaAs quantum dots \cite{shulman2012}.

Electrons confined in interface quantum dots could also be physically shuttled along the interface. This would open up the possibility of transporting spin quantum information that might have been encoded and manipulated while the electron was bound elsewhere. A recent proposal discusses the use of a charge-coupled device (CCD) structure to realize a surface-code quantum error correction architecture \cite{pica2016surface}. The surface code requires a two-dimensional lattice of data and measurement qubits \cite{fowler2012}. Ref.~\cite{pica2016surface} envisages using interface dots as the data qubits, and donor electrons as the measurement qubits. SWAP operations between donors and dots, combined with electron shuttling across the array controlled by the CCD, provide the necessary functions. 

This scheme also requires that the spin quantum state is unaffected by the electron shuttling operation. The spin-preserving, coherent shuttling of an electron spin at the Si/SiO$_2$ interface has been recently demonstrated \cite{hensen2020silicon}, using a CMOS device architecture identical to those adopted for the implanted-donor devices studied in our group.

A long-distance coupling method similar to the spin chain can be achieved with an elongated, multi-electron quantum dot. A dot with an odd electron number typically behaves just like an odd-number spin chain, albeit with a less trivial gap to the excited states. 

Interestingly, a dot with even electron number can also mediate a long-distance interaction between qubits coupled at its ends, via a form of Ruderman-Kittel-Kasuda-Yosida (RKKY) interaction \cite{srinivasa2015} that involves a virtual tunneling between the electron spin qubits and the dot. The ground state of the dot is a $S=0$ singlet and, contrary to the case of the odd spin chain, the gap between the ground and the higher excited states needs not be particularly large; its value can be used to control the effective coupling strength between qubits coupled to the dot. 

Experimental demonstrations of exchange coupling between distant spins via a mediator quantum dot have been achieved in GaAs/AlGaAs systems, both in the many-electron regime \cite{malinowski2019} and in the curious case where the mediator dot is empty of electrons \cite{baart2017}. Similar experiments are currently underway in silicon, including with donors. 

\subsection{Coupling via ferromagnets}

A ferromagnetic material with a sizable magnon gap can be used to mediate an effective coupling between spin qubits that are dipolarly coupled to it \cite{trifunovic2013}. The coupling strength can be adjusted by making the spin qubits energy splitting approach the magnon gap of the ferromagnet. For a pair of spins placed 25~nm below the ferromagnet at a mutual distance of 1~$\mu$m, a 2-qubit SWAP gate can be achieved in $\sim 10 - 100$~ns. Because this method does not involve any control of charge states, nor an atomically-precise placement of the spins with respect to the ferromagnet, it is particularly suitable for implanted donor qubits or, equivalently, for nitrogen-vacancy centres in diamond, where a coherent interaction between nanodiamonds and the spin waves in yttrium-iron garnet has been demonstrated \cite{andrich2017}.

\subsection{Coupling via microwave photons}

The largest distance between donor spin qubits on a chip can be achieved using the paradigm of cQED \cite{schoelkopf2008}, where a quantum system is coherently coupled to a microwave photon confined to a superconducting cavity. Qubits at opposite ends of the cavity can be made to interact and perform 2-qubit logic gates while being spaced at distances $\sim 1$~cm apart, corresponding to the half-wavelength of photons in the $5-10$~GHz range. 

Magnetic coupling of photons to spin ensembles is described in Section~\ref{sec:ensemble_devices}. Extending cQED methods to individual spins requires reaching the regime where the coupling of a single donor to the cavity, $g_0$, exceeds both the photon decay rate $\kappa$ and the spin dephasing rate $\Gamma$. We have quantitatively analyzed the direct coupling between a donor spin and the vacuum magnetic field of a cavity, and estimated that an optimized cavity design can yield $g_0 \approx 3$~kHz \cite{tosi2014}. This value is very low compared to the $\sim 10$~MHz couplings easily obtained in superconducting qubits, but it should be compared to the intrinsic linewidth of donor spin qubits in enriched $^{28}$Si, which has been observed to be as low as $\Gamma = 1.8$~kHz \cite{muhonen2014}. The next condition to achieve strong coupling, $\kappa < 3$~kHz, would require cavities with quality factors exceeding $10^6$, which is an achievable goal \cite{megrant2012}, albeit made more challenging in the presence of the strong ($\sim 0.2$~T) magnetic field necessary to bring the spin energy splitting in resonance with the cavity photons. 

\begin{figure}[tb]
  \includegraphics[width=8.2cm]{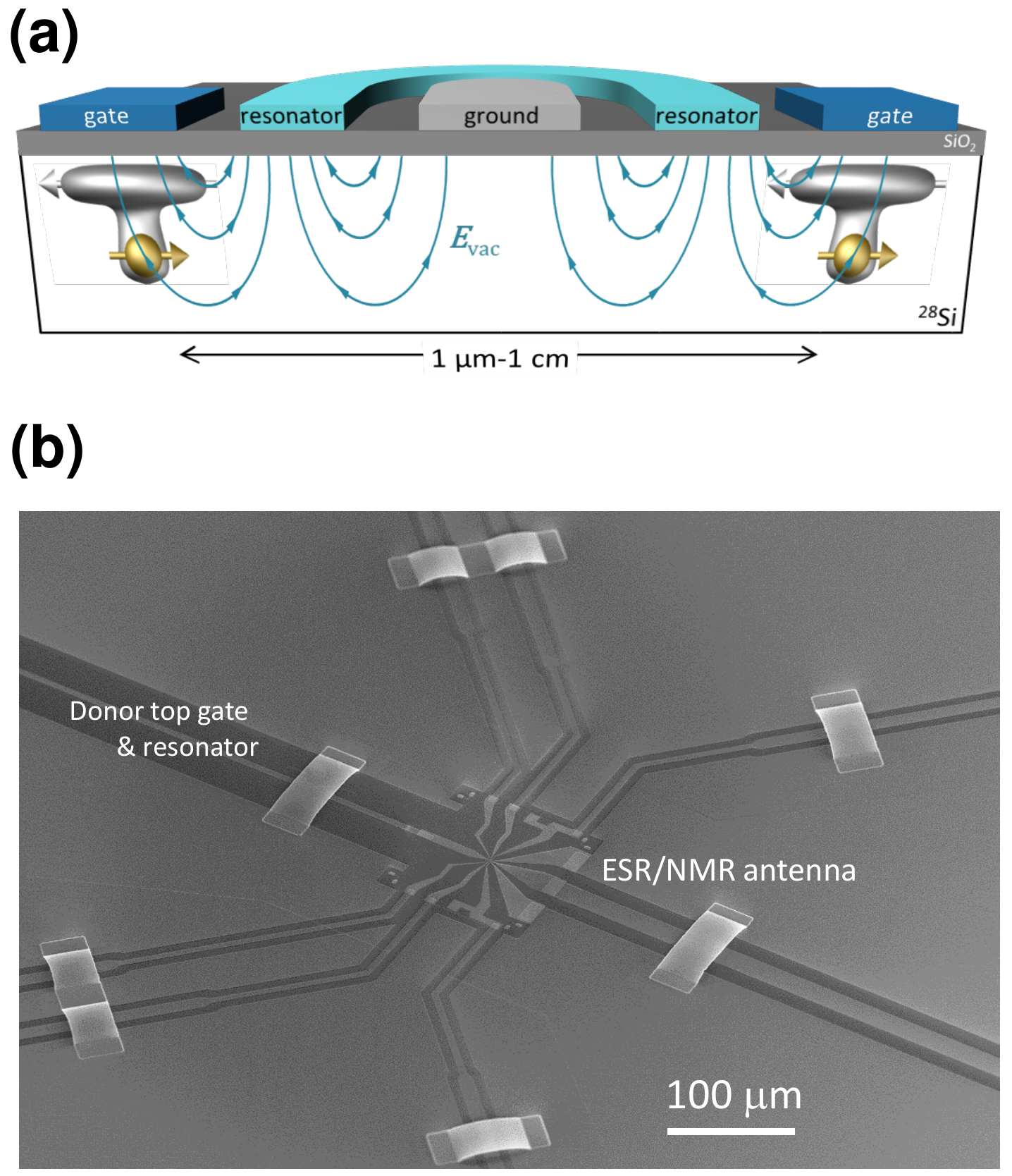}
  \caption{Coupling flip-flop qubits to microwave photons. (a) Sketch of long-distance coupling between flip-flop qubits, mediated by the vacuum electric field $E_{\rm vac}$ in a microwave resonator (adapted from \cite{tosi2017}). (b) Scanning electron micrograph of a flip-flop qubit device, integrated with a superconducting microwave resonator. 
  \label{fig:ff_cavity}
  }
\end{figure}

To circumvent these challenges, we have proposed \emph{electrically} coupling single-donor flip-flop qubits via microwave photons, exploiting the large electric dipole that can be induced on the donor by pulling the donor-bound electron away from the nucleus [Figure~\ref{fig:ff_cavity}(a)]. In this case, we predict an electric flip-flop / photon coupling strength $g_{\rm E}^{\rm ff} \sim 3$~MHz \cite{tosi2014}, i.e. 3 orders of magnitude larger than the magnetic coupling to the spin. This makes the flip-flop qubit an ideal system to couple to high kinetic inductance resonators, where the vacuum electric field is maximized, and which can be fabricated to achieve sufficiently high quality factors even in the presence of strong magnetic fields \cite{samkharadze2016}.

An intriguing prospect enabled by the flip-flop qubit idea is the coupling of a single \emph{nuclear} spin to a microwave photon. This could be achieved via a Raman-like transition where a (classical) magnetic ac field drives the $\ket{\downarrow\Downarrow} \leftrightarrow \ket{\uparrow\Downarrow}$ electron spin transition, while the cavity photon electrically drives the $\ket{\downarrow\Uparrow} \leftrightarrow \ket{\uparrow\Downarrow}$ flip-flop one. This results in an effective nuclear-photon coupling of order 1~MHz \cite{tosi2018} -- a striking outcome, unattainable with other nuclear spin systems. 

Our current experimental focus is on developing fabrication processes for high-quality superconducting cavities, compatible with the process flow used for the fabrication of MOS single-donor spin qubits [Figure~\ref{fig:ff_cavity}(b)]

\section{High-spin nuclei: quadrupole interactions and its applications}

\subsection{Origin and detection of nuclear quadrupole splitting }  \label{sec:quadrupole}
Nuclei with a spin $I > 1/2$ can exhibit a non-spherical charge distribution which is described by a quadrupole moment (see Table I). The charge distribution has an axis of symmetry that aligns with the nuclear angular momentum and interacts with an electric field gradient (EFG) $V_{\rm\alpha\beta}$ (where $\alpha$ and $\beta$ are principal axes in the local crystal coordinate system) produced by external charges. The interaction is represented by the following quadrupole Hamiltonian (in frequency units):
\begin{equation}
\label{eq:QuadHam}
H_{\rm Q} = \gamma_{\rm S}\frac{e\mathcal{Q}V_{\rm zz}}{4I\left(2I-1\right)h}\left[3I^2_{\rm z}-\mathbf{I}^2+\eta\left(I^2_{\rm x}-I^2_{\rm y}\right) \right] 
\end{equation}\\
where $\gamma_{\rm S}$ is a multiplicative scaling factor (resulting from the Sternheimer anti-shielding effect \cite{kaufmann1979}), $e$ is the electron charge, $h$ is Planck's constant, $\mathbf{I}$ is the nuclear spin operator with components $I_{\rm \alpha}$, $I$ in the denominator is the scalar value of the nuclear spin (e.g. $I = 9/2$ for $^{209}$Bi) and $\eta = \left(V_{\rm xx} - V_{\rm yy}\right)/V_{\rm zz}$ is an asymmetry parameter.

The charge distribution of the donor-bound electron is a potential source of EFG and thus quadrupole interaction. Silicon has a conduction band minimum that is six-fold degenerate along the $\left<100\right>$ equivalent crystallographic directions --- commonly referred to as ``valleys'' \cite{ando1982}. The degeneracy of these valleys is broken by the confining potential of the donor, resulting in a singlet A$_{\rm 1}$ ground state and doublet E and triplet T$_{\rm 1}$ excited states \cite{kohn1955}. The A$_{\rm 1}$ ground state of the donor electron comprises an equal combination of Bloch wavefunctions at each of the valleys, providing a high degree of symmetry for this state. Evaluating the EFG tensor components $V_{\rm\alpha\beta} = \langle\psi\vert\frac{\partial^2 V(\textbf{r})}{\partial\alpha\partial\beta}\vert\psi\rangle$, where $V(\textbf{r})$ is the electrostatic potential produced at the nucleus by the electron charge with a wavefunction $\vert\psi\rangle$, gives $V_{\rm\alpha\beta} = 0$ for the symmetric case of $\vert\psi\rangle = \vert A_{\rm 1}\rangle$. For a donor in a bulk silicon crystal (in the absence of strain and electric fields) the electron is perfectly described by the singlet ground state $\vert\psi\rangle = \vert A_{\rm 1}\rangle$ and the quadrupole interaction is therefore non-existent. It is through mixing the donor ground state wavefunction with the excited states that a non-vanishing EFG tensor is realized.

Mixing of the electron ground and valley excited states can occur in the presence of strain, electric fields and defects. It has been suggested that ion-implantation-induced defects at the Si/SiO$_2$ interface can modify the wavefunction of shallow donors (within a few nanometers of the defects) and induce shifts on the donor NMR frequencies through the quadrupole interaction \cite{ mortemousque2016}. Experiments have also reported the observation of quadrupole shifts on the NMR frequencies of $^{75}$As donors resulting from strain intentionally introduced to the silicon through differential thermal contraction \cite{franke2016}. 

Electric field gradients may be present in a device even when the donor electron is absent. Strain distorts the positions of the silicon atoms coordinating the silicon-donor covalent bonds, reducing the symmetry of the donor site and producing an EFG \cite{asaad2020}. Quadrupole interactions have been observed in the NMR spectra of ionized $^{75}$As nuclear spin ensembles \cite{franke2015}, as well as in a single $^{123}$Sb nucleus in a nanodevice \cite{asaad2020}.

\subsection{Nuclear electric resonance}  \label{sec:NER}

In 1961, Bloembergen predicted that a nucleus or a paramagnetic ion with spin $>1/2$ placed in a lattice site that lacks point inversion symmetry could be controlled via oscillating electric fields \cite{bloembergen1961}, as opposed to the standard magnetic resonance methods. In the case of nuclei, this arises because an electric field can modulate the nuclear quadrupole interaction.

We have discovered Nuclear Electric Resonance (NER) in an ionized $^{123}$Sb$^+$ donor (spin $I=7/2$) \cite{asaad2020}, integrated with our standard spin qubit device structure on a $^{28}$Si epilayer (Figure~\ref{fig:donor_device}). The oscillating electric field at frequencies $\approx 8$~MHz is provided by the same metallic gate used to tune the donor electrochemical potential. A static quadrupole splitting $\nu_Q \approx 66$~kHz separates all the nuclear resonances and allows their individual addressing. The electrical control of an ionized donor is significant because the absence of the hyperfine-coupled electron preserves the exceptionally long dephasing times ($T_{\rm 2n+}^* \approx 100$~ms in this device), in contrast to other methods of electrical control based on the modulation of the hyperfine coupling tensor \cite{thiele2014,sigillito2017,morello2017}.

Nuclear spin transitions that change the spin projection by one quantum of angular momentum, $\Delta m_I = \pm 1$, are induced at a rate proportional to $\vert \bra{m_I-1}I_xI_z + I_zI_x\ket{m_I}\vert$ by the quadrupole interaction, which is notably zero for the $-1/2 \leftrightarrow +1/2$ transition. However, the nuclear quadrupole interaction is quadratic in the spin operators (Eq.~\ref{eq:QuadHam}), and allows $\Delta m_I = \pm 2$ transitions that can be used to access all the nuclear spin projections.

We studied the microscopic mechanism that enables NER, and found a remarkable quantitative agreement between the data and a model that describes the electric field gradient modulation as arising from a local distortion of the charges in the atomic bonds between the $^{123}$Sb donor and its neighbouring Si atoms. The static quadrupole splitting, $\nu_Q \approx 66$~kHz, is well described by the effect of local strain caused by the different thermal expansion coefficients of the silicon host and the aluminum gates fabricated on top of it. Therefore, this experiment helps to validate quantitative models that link strain to nuclear resonance shifts (Section~\ref{sec:strain_sensing}).

\begin{figure}[tb]
  \includegraphics[width=8.2cm]{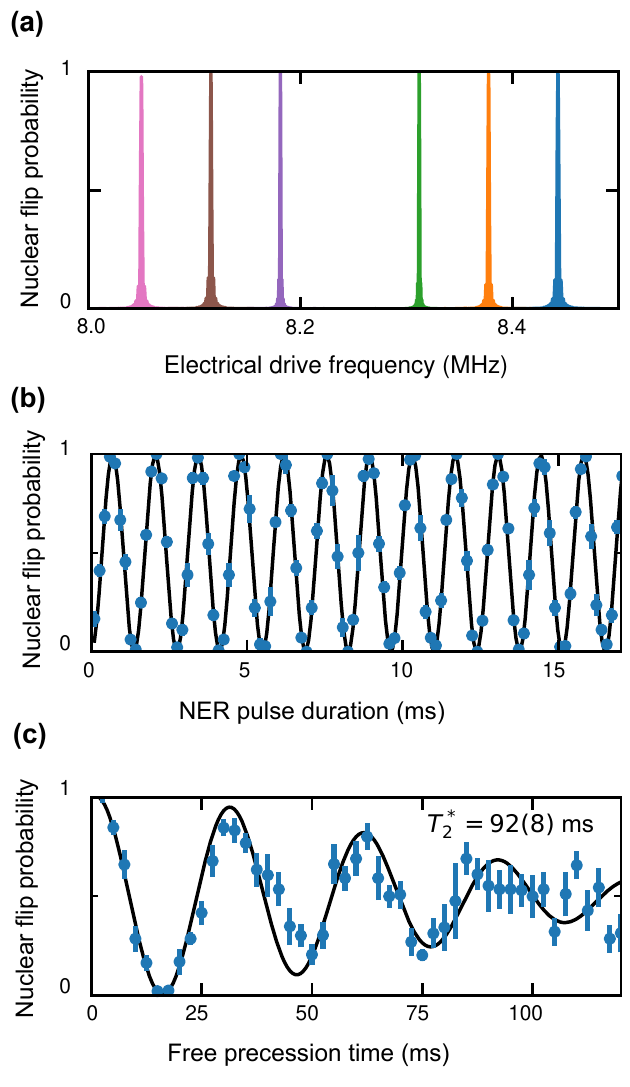}
  \caption{Nuclear electric resonance in a single ionized $^{123}$Sb$^+$ nucleus. (a) NER spectrum for the $\Delta m_I = \pm 1$ transitions. Notice the static quadrupole splitting $\nu_Q \approx 66$~kHz between the resonances, and the absence of the $-1/2 \leftrightarrow +1/2$ transition. (b) Electrically-driven Rabi oscillation on the $+5/2 \leftrightarrow +7/2$ transitions, and (c) electrically-driven Ramsey experiment to extract the dephasing rate $T_{2n+}^*=92(8)$~ms. Adapted from \cite{asaad2020}.
  \label{fig:NER}
  }
\end{figure}

\subsection{Quantum chaos}  \label{sec:chaos}

The observation of a sizable quadrupole splitting in a single, highly coherent, high-spin nucleus \cite{asaad2020} opens the path to the experimental demonstration of a single chaotic ``kicked-top''.

The kicked top has been among the most studied model systems for quantum chaos for many decades \cite{haake2000}. It has been experimentally realized in cold atomic ensembles \cite{chaudhury2009}, but never in a single quantum system. We have theoretically proposed, and quantitatively analyzed, the embodiment of a periodically-driven top (a more experimentally feasible variant of the kicked top, where the system is subjected to a sinusoidal drive instead of a train of $\delta$-functions) using a the nuclear spin of a group-V donor in silicon \cite{mourik2018}. 

The classical Hamiltonian of a spinning top with angular momentum $\mathbf{L}=\left(L_\mathrm{x} \hspace{0.5em} L_\mathrm{y} \hspace{0.5em} L_\mathrm{z}\right)^\mathrm{T}$ ($|\mathbf{L}|=$ constant), subjected to a periodic drive at frequency $\nu$, has the form:
\begin{equation} \label{eq: H_top_classical}
\mathcal{H}_\mathrm{classical} = c_1 L_\mathrm{z}+ c_2 L_\mathrm{x}^2+ c_d \cos \left(2 \pi \nu t \right) L_\mathrm{y}.
\end{equation}
The key term is the quadratic one, $c_2 L_\mathrm{x}^2$, which makes the precession frequency a nonlinear function of the orientation of the top.

It is easy to see that the quantum equivalent of such Hamiltonian can be realized using a nucleus with spin $I>1/2$ described by:
\begin{equation} \label{eq: H_top_quantum}
\mathcal{H}_\mathrm{quantum} = \gamma_\mathrm{n}B_0  I_\mathrm{z} + Q I_\mathrm{x}^2 + \gamma_\mathrm{n} B_1 \cos \left( 2 \pi \nu t \right) I_\mathrm{y},
\end{equation}  
where the quadratic term, $Q I_\mathrm{x}^2$, is provided by the nuclear quadrupole coupling. We predict that, within the range of experimentally accessible parameters, it will be possible to explore the transition from regular to chaotic dynamics in such coherent, controllable and QND-measurable system \cite{mourik2018}. This research is linked to the fundamental question of the nature of the quantum-classical transition \cite{zurek1994}, and the thermalization of isolated quantum systems \cite{rigol2008}. More recently it was understood that the proliferation of errors in digital quantum simulation can be mapped to the emergence of chaos in a kicked top \cite{sieberer2019}, providing a profound link between quantum chaos and quantum information processing. Experiments are currently underway to observe chaotic dynamics in a single-nucleus periodically-driven top, as described in \cite{mourik2018}, using a single ion-implanted $^{123}$Sb nucleus in silicon \cite{asaad2020}.

\begin{figure}[tb]
  \includegraphics[width=8.2cm]{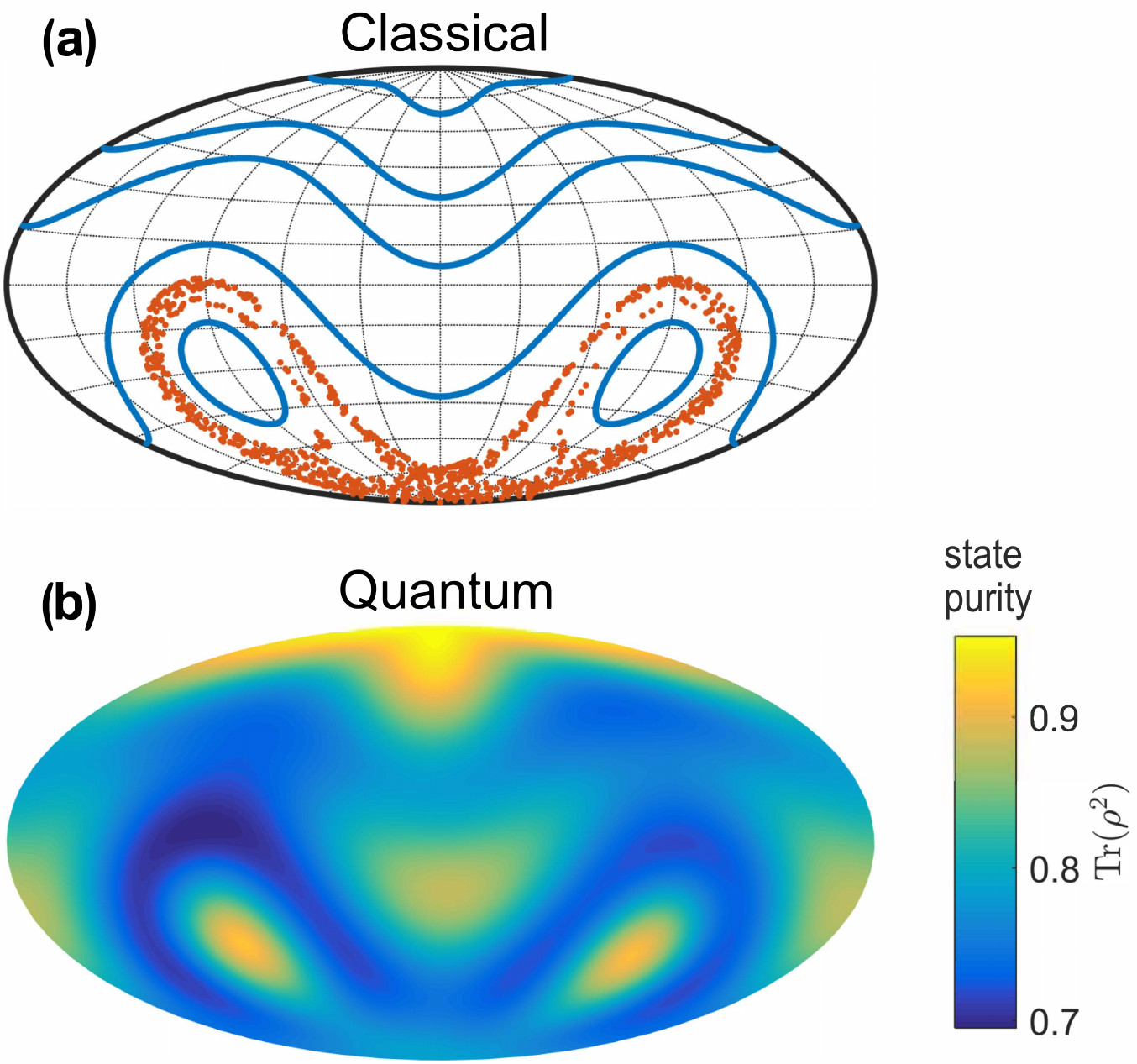}
  \caption{Quantum chaos in a single high-spin nucleus. (a) Simulated Poincar{\'e} map of a classical periodically driven top (Eq.~\ref{eq: H_top_classical}): the blue lines described closed, regular trajectories, while the orange dots display the stroboscopic location of the top when prepared in a chaotic region. (b) Quantum state purity of a nuclear spin described by the Hamiltonian in Eq.~(\ref{eq: H_top_quantum}), with parameters ratios identical to those used for the classical simulation. In addition, the quantum system is subjected to incoherent noise during the evolution. The state purity is most significantly reduced in the same regions that display dynamical chaos in the classical system. Adapted from \cite{mourik2018}.
  \label{fig:chaos}
  }
\end{figure}

\section{\label{sec:ensembles}Quantum memory with small donor ensembles}

\begin{figure}[tbph]
  \includegraphics[width=8.2cm]{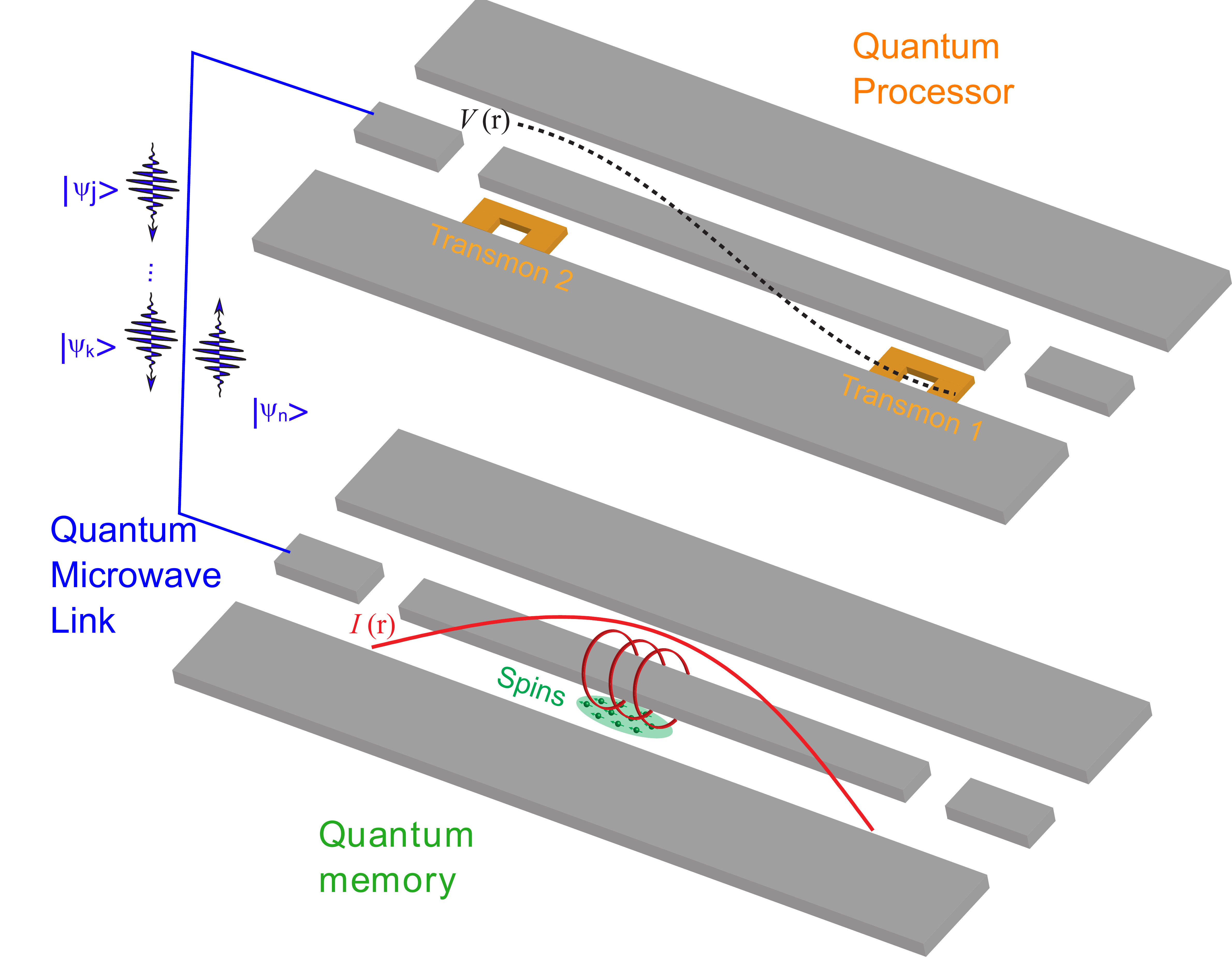}
  \caption{Quantum Turing Machine principle. A small-scale transmon-based Quantum Processor (shown here with 2 qubits) is interfaced with a spin-ensemble Quantum Memory via a quantum microwave link, enabling the high-fidelity exchange of quantum states. 
  \label{fig:QMscheme}
  }
\end{figure}

Using donors in silicon as qubits requires individual addressing of the spins as well as a way to couple them together, as described in the previous sections. Another possible direction for donor-based quantum technologies would be to use instead ensembles of donor spins as a quantum memory, able to store and deterministically retrieve quantum states originating from a superconducting quantum processor. The motivation here is to develop a quantum Turing machine, consisting of (i) a small-scale superconducting quantum processor (QP) with $N_{QP}=2-10$ individually addressable transmon qubits, (ii) a quantum memory (QM) based on an ensemble of typically $N=10^5 - 10^7$ donor spins and (iii) a microwave photon link between the two modules. Qubit states are initialized, processed and read-out by the QP, but are stored in collective degrees of freedom of the spin ensemble (spin-waves) \cite{wesenberg2009}. No individual addressing of the spins is necessary; instead, one needs a high-fidelity quantum interface between microwave photons and spin-waves. A schematic description is shown in Figure~\ref{fig:QMscheme}.

Let us briefly discuss the interest of such a novel universal quantum computing architecture. A spin-ensemble QM can in principle store many quantum states in parallel, with an upper bound given by $N$. Provided a high-fidelity quantum microwave photon link can be implemented between the QP and the QM, this Quantum Turing Machine therefore possesses a large number of qubits (of order $N$), while necessitating individual addressing of only $N_{QP}$ qubits, many orders of magnitude smaller than $N$, thus overcoming one of the major issues of existing quantum computing schemes. Moreover, the cross-talk would be intrinsically very low since data or ancilla qubits would be stored in a separate device during quantum gate operation. On the other hand, the fault-tolerant aspect and quantum error correction capability of this scheme needs to be theoretically addressed.

High-fidelity interfaces between a QP and a microwave photonic link have already been demonstrated in several experiments \cite{wenner_catching_2014} using a circuit QED architecture in which the qubits are coupled to a superconducting cavity (see Figure~\ref{fig:QMscheme}); thus the main challenge lies in the storage of incoming microwave photons prepared in a state $\ket{\psi_j}$ into the QM, and in their deterministic retrieval in a quantum state as close as possible to $\ket{\psi_j}$. Here again, cQED provides a natural method to interface incoming microwave photons with a spin-ensemble through a superconducting resonator, as seen in Figure~\ref{fig:QMscheme}. Because the microwave field needs to be in its ground state at thermal equilibrium and the spin ensemble fully polarized, natural operating conditions would be a frequency around $5-8$\,GHz and a temperature of $10$\,mK.

Experimental efforts towards the implementation of a spin-ensemble quantum memory have been so far focused on NV centers in diamond, and were recently reviewed \cite{grezes_towards_2016}. Efficient absorption of an incoming microwave photon requires the spin ensemble to be coupled to the cavity in the high cooperativity regime; this was shown to be possible in several spin systems \cite{kubo2010,schuster_high-cooperativity_2010,probst_anisotropic_2013}, including donors in silicon \cite{rose_coherent_2017}. Retrieving deterministically the desired state is more demanding. A simple two-pulse echo sequence is not sufficient, as the noise from spontaneous emission would forbid high-fidelity state retrieval. Proposed schemes, based on theoretical and experimental work in optical quantum memories \cite{damon2011}, involve two consecutive $\pi$ pulses, together with a mechanism to `silence' the first echo, for instance by dynamic detuning of the resonator \cite{afzelius2013,julsgaard2013}. During its operation the QM will at times need to be reset to its ground state. This can be achieved either by optical pumping or by letting the electron spins relax by spontaneous emission of a microwave photon into the cavity mode, via the Purcell effect \cite{bienfait2016}.

The most advanced experiment demonstrated the storage of a microwave field containing 1 photon on average in a NV spin ensemble in diamond, and its retrieval after a $100~\mu\mathrm{s}$ storage time \cite{grezes2015}. The motivation to puruse this scheme with donors in silicon is their much longer coherence time, opening the way to much longer memory storage times. Coherence times of order 10~ms are readily achievable \cite{bienfait2015}. Bismuth donors in silicon stand out as particularly interesting in that context. First, several ESR transitions are accessible at low field around $7.4$\,GHz, due to the large hyperfine splitting and the $I=9/2$ nuclear spin of the $^{209}$Bi atom \cite{mohammady_bismuth_2010}. Moreover, bismuth donors can be biased at the so-called clock transitions, where the coherence time was shown to reach as long as 2 seconds in bulk silicon \cite{wolfowicz_atomic_2013}. 

\section{Quantum sensing} \label{sec:sensing}

\subsection{Magnetic sensing with single donors} \label{sec:magnetic_sensing}

A qubit system can be used as a quantum sensor by exploiting the dependence of the precession frequency on some external (usually classical) parameter that appears as a longitudinal term in the qubit Hamiltonian. For a spin qubit, the natural longitudinal term is the magnetic Zeeman energy $\epsilon_Z = \gamma B_z$ (see Eq.~\ref{eq: H_donor}, where the gyromagnetic ratio $\gamma$ has units of Hz/T). 

A small perturbing dc magnetic field $b$ added to the static field $B_0$ can be detected by performing a Ramsey experiment on the spin: the extra field induces a phase accumulation $\phi = 2 \pi \gamma b \tau_{\rm i}$ over an interrogation time $\tau_{\rm i}$. The optimal sensitivity is obtained by using $\tau_{\rm i} \approx T_2^*$, and takes the value \cite{taylor2008}:
\begin{equation}
    \eta_{\rm dc} \approx \frac{1}{2\pi\gamma C \sqrt{T_2^*}}
\end{equation}
where $C$ is a factor that accounts for the detection efficiency of the spin. For both the electron and the nuclear spin of single donors detected in single-shot mode \cite{morello2010,pla2013}, $C \equiv \mathcal{F}_{\rm meas} \approx 1$. Inserting the values of $T_2^*$ measured on a single P donor in enriched $^{28}$Si \cite{muhonen2014} (see Table~\ref{tab:benchmarks}), we find $\eta_{\rm dc} \approx 0.3$~nT/$\sqrt{\mathrm{Hz}}$ for the electron spin, and $\approx 10$~nT/$\sqrt{\mathrm{Hz}}$ for the ionized nucleus. 

Much improved sensitivities can be obtained by adopting some form of dynamical decoupling like CPMG. The application of a train of refocusing $\pi$-pulses separated by a time interval $\tau$  removes the effect of static disturbances, and effectively makes the spin sensitive only to ac  magnetic fields with frequencies centered around the passband frequency $\nu_{\rm p} = 1/2\tau$ \cite{cywinski2008,muhonen2014}. The ac magnetic field sensitivity becomes optimal for signals with spectrum centered around $1/T_2^{\rm CPMG}$ and takes the value \cite{taylor2008}:
\begin{equation}
    \eta_{\rm ac} \approx \frac{1}{4\gamma  \sqrt{T_2^{\rm CPMG}}}
\end{equation}
Using again the values from Ref.~\cite{muhonen2014} and Table~\ref{tab:benchmarks}, we find $\eta_{\rm ac} \approx 10$~pT/$\sqrt{\mathrm{Hz}}$ for the electron spin, and $\approx 2$~nT/$\sqrt{\mathrm{Hz}}$ for the ionized nucleus. The magnetic field sensitivity can be measured directly via a noise spectroscopy experiment, which on the P donor electron indeed revealed a noise floor equivalent to $\eta_{\rm ac} \approx 14$~pT/$\sqrt{\mathrm{Hz}}$ \cite{muhonen2014}.

These values are extremely competitive for atomic-scale sensors in the solid state, owing to the exceptional coherence times of donor spins. Notably, such coherence times were obtained with donors read out electrically via spin-to-charge conversion (Section~\ref{sec:single_donors}), and therefore placed within $\sim10$~nm from the Si/SiO$_2$ interface \cite{mohiyaddin2013}. This means that these sensitivity values describe a spin sensor placed $\sim20$~nm beneath the surface of the chip, on which some object or material of interest might be deposited.

An interesting future direction to further enhance the sensitivity, made possible by the ability to control high-spin nuclei like $^{123}$Sb \cite{asaad2020}, is the use of non-classical spin states such as NOON or squeezed states \cite{pezze2018}, or even exploiting the chaotic dynamics of a periodically-driven high-spin nucleus \cite{fiderer2018}.

\subsection{Strain sensing} \label{sec:strain_sensing} 
Donors in silicon exhibit a sensitivity to strain through the dependence of the electron g-factor, hyperfine constant, and/or quadrupole interaction on lattice deformation. For small strains ($\vert\varepsilon\vert<{10}^{-3}$), the donor spin transition frequencies exhibit a strong linear dependence on the hydrostatic component of an applied strain $\varepsilon_{hs}$ via the hyperfine constant $A$ \cite{mansir2018}. The measured linear coefficient $K$ relating strain to a change in the hyperfine coupling $\Delta A/A = K\varepsilon_{hs}$ for each group-V donor is displayed in Table~\ref{tab:linearhyperfine}.

The strain sensitivity of donors in silicon might be utilized in a range of applications, from quantum computing to hybrid donor-mechanical systems and metrology. For $^{31}$P, a strain of order $\varepsilon \sim 10^{-4}$ would generate a change in the hyperfine constant of $\Delta A = 1$~MHz and an equivalent spin resonance shift of $\Delta \nu = (d\nu/dA)\Delta A \approx  0.5$~MHz (where $d\nu/dA$ is the sensitivity of a given transition to $A$), a value far exceeding typical donor spin linewidths in isotopically-enriched $^{28}$Si ($\sim 2$~kHz \cite{muhonen2014}). Such a strain can be generated with piezoelectric materials on the surface of the silicon chip \cite{dreher2011}. This effect could be used to actuate quantum logic operations in a donor-based quantum processor, where strain is applied by local piezoelectric gates to bringing individual qubits in and out of resonance with global microwave and radio frequency control fields \cite{dreher2011}.

\begin{figure}[tbph]
  \includegraphics[width=7.8cm]{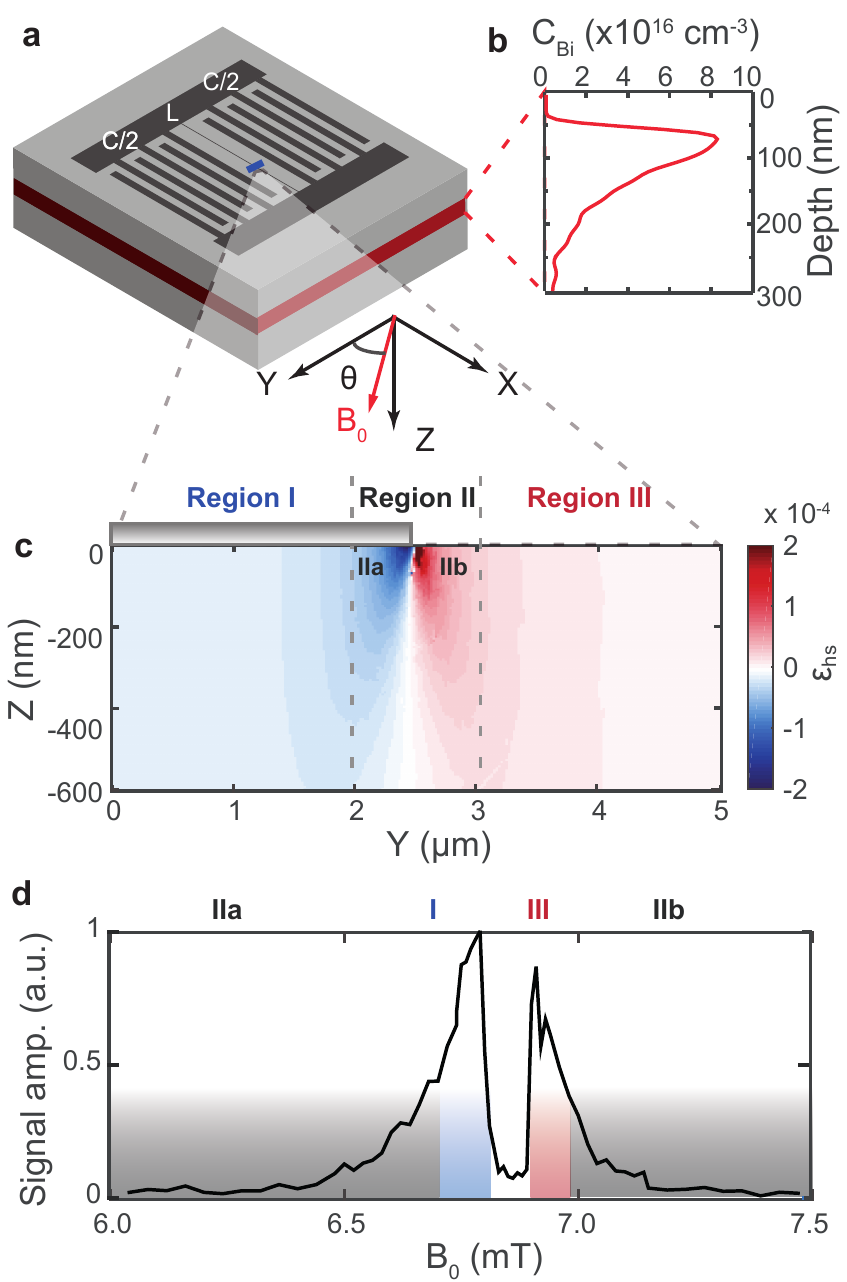}
  \caption{\label{fig:microres} (a) Superconducting LC micro-resonator fabricated on a bismuth-doped $^{28}$Si sample. (b) Bismuth ion implantation profile measured using secondary ion mass spectroscopy. (c) Finite element simulation of the hydrostatic strain $\varepsilon_{hs} = (\varepsilon_{xx}+\varepsilon_{yy}+\varepsilon_{zz})/3$ in the sample resulting from differential thermal contraction of the silicon and aluminum. (d) ESR spectrum measured through the LC resonator for a $^{209}$Bi:Si spin transition. The transition shows a single symmetric peak in bulk experiments but here is split and asymmetric. The splitting is caused by the hydrostatic strain in (c), which is compressive underneath the wire and tensile to the side, generating negative and positive hyperfine shifts, respectively, for donors in these regions. The spectrum has been coloured and labelled according to where in the device each segment originates.
  }
\end{figure}

Single donors in isotopically-enriched $^{28}$Si substrates could be used to detect strains down to ${10}^{-8}$ ($^{209}$Bi) by measuring shifts in spin resonance frequencies. These single-atom sensors could help map strain variations in silicon-based quantum devices. ESR experiments on small ensembles of bismuth donors underneath high-sensitivity aluminum micro-resonators has already shown promise in this direction \cite{pla2018}. Strain due to thermal contraction of the aluminum ESR circuit produced a donor ensemble spin resonance spectrum with non-bulk-like features (see Figure~\ref{fig:microres}). Segments of the spectrum could be traced to donors in specific regions of the device; for example, the highly detuned tails of the resonance spectra were attributed to donors at the edge of the circuit, which experienced the greatest strain. The correlation between resonance frequency and donor location in these micro-resonator studies might be exploited to investigate decoherence mechanisms in silicon devices, where the coherence time of donors residing in different parts of the circuit could be probed through the use of frequency-selective pulse sequences.

For high nuclear spin donors, the quadrupole interaction can also be used to measure device strains. Spin linewidths of $\sim 2$~Hz measured for a single $^{123}$Sb$^+$ nucleus translate into a strain sensitivity of $10^{-8}$ \cite{asaad2020}, the same order of magnitude as detectable through the hyperfine interaction.

\begin{table}
\begin{tabular}{cccc}
\hline
Donor & $K$ & $\vert d\nu/dA \vert$ (low field) & $\vert d\nu/dA \vert$ (high field)\\
\hline \\
$^{31}$P & $79.2$ & 1 & 0.5 \\
$^{75}$As & $37.4$ & 2 & 1.5 \\
$^{121}$Sb & $32.8$ & 3 & 2.5 \\
$^{209}$Bi & $19.1$ & 5 & 4.5\\
\hline
\end{tabular}
\caption{\label{tab:linearhyperfine} Linear coefficient $K$ (in units of inverse strain) of the second-order hyperfine strain model \cite{mansir2018} for the group-V donors in silicon. Also listed is the maximum hyperfine sensitivity among the donor spin transitions both in the low magnetic field limit ($\gamma_e B_0 \ll A$) and the high field limit ($\gamma_e B_0 \gg A$).}
\end{table}

\section{\label{sec:implantation}Scale-up technologies: deterministic single-ion implantation}

Exploiting the promising physical attributes of donor spins in semiconductors requires a strategy for constructing a large-scale device with engineered features at the nanoscale. This strategy will include a device architecture tolerant of finite manufacturing precision, a fabrication technique ideally based on standard industry processes \cite{rubin2003} and the versatility to incorporate different dopant atoms into the device. As introduced in section~\ref{sec:single_donors}, ion implantation meets these requirements. However, a method is required to overcome Poisson variations in donor number caused by the stochastic nature of the ion source \cite{jamieson2017}.  Methods in use or under development include single ion traps \cite{groot2019}, single ion flight detectors \cite{racke2018}, secondary electrons induced from single ion impacts \cite{schenkel2015}, and transistor transients \cite{johnson2010}. 

A method compatible with the process-flow for the fabrication of large-scale devices employs the signal generated by the induction of charge on biased surface detector electrodes following the dissipation of kinetic energy by ionization after ion impact \cite{vandonkelaar2015}.  This signal registers the implantation of a single ion, where the implant site is localized with a nanostencil fabricated in an AFM cantilever \cite{persaud2005,meijer2008}.  The signal can be used to reposition the nanostencil to the next implant site to construct a large-scale device.  Post-implantation annealing at 900 $^\circ$C for 10~s is required to activate the $^{31}$P donors. 

The use of on-chip single ion detectors was already demonstrated in 2007, when the time-resolved control of the charge state of a donor pair was achieved using a device fabricated with precisely two counted $^{31}$P ions \cite{andresen2007}. A similar method has also been adopted in the semiconductor foundry at Sandia National Laboratories, where on-chip ion detectors were successfully integrated within a CMOS process flow \cite{singh2016}, and a precise placement of the ions was obtained using a focused ion beam \cite{pacheco2017}.

\begin{figure}[tbph]
  \includegraphics[width=8.2cm]{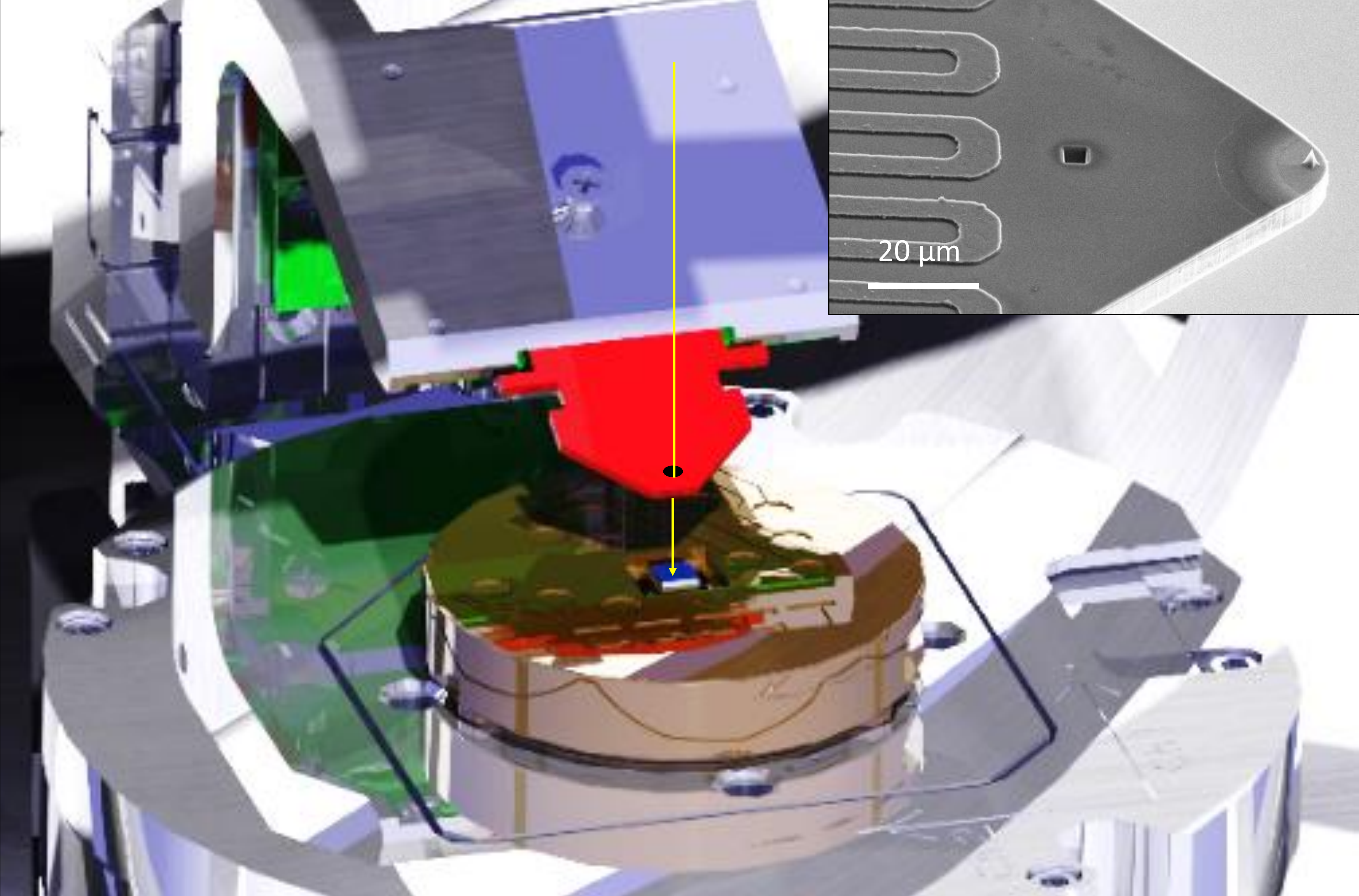}
  \caption{\label{fig:stencil} Illustration of the scanned nanostencil concept showing a cantilever (red) with a nanoscale aperture used to direct an ion beam (yellow) to a specific location on the sample (dark blue).  The inset shows the nanostencil in the cantilever, where an aperture (not shown) within the central milled square trench collimates the beam.
  }
\end{figure}

\begin{figure}[tbph]
  \includegraphics[width=8.2cm]{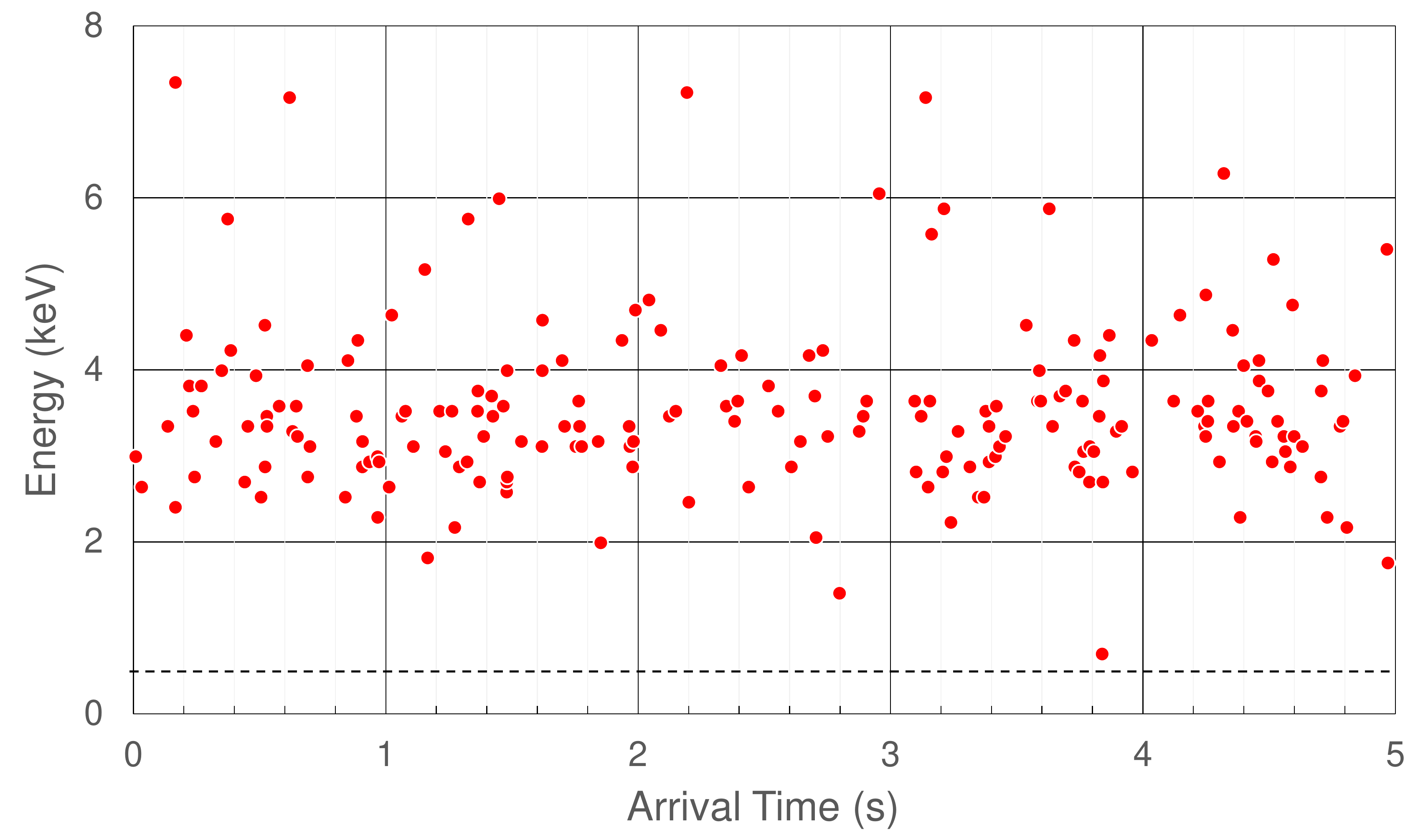}
  \caption{\label{fig:pulse_height} Experimental pulse height of two hundred 14 keV P$^+$ ions incident on a silicon device, measured with on-chip electrodes and optimized charge-sensitive electronics.  The arrival time is stochastic, and the scatter in the detected energy is due to ion straggling and Poisson statistics. The dashed line represents the noise threshold of the charge-sensitive electronics.
  }
\end{figure}

For $^{31}$P$^+$ ions implanted 20 nm deep in silicon, sufficiently close to the surface to allow control by electrostatic gates (Figure~\ref{fig:donor_device}), the incident kinetic energy should be 14 keV, of which $\sim$~3.5 keV results in ionization in the substrate corresponding to the generation of $\sim$~1000 electron-hole (e-h) pairs.  The missing kinetic energy is lost to the surface oxide dead layer, nuclear stopping (phonons), and recombination in the plume of charge along the ion track that partially shields the bias field from the electrodes.  Highly sensitive low-noise external preamplifier circuitry is necessary to detect the e-h pair signal from the on-chip electrodes: the present state-of-the-art is a noise threshold of $\approx 400$~eV equivalent to $\approx 110$~e-h pairs.  Experimental data for two hundred consecutive ions is shown in Figure~\ref{fig:pulse_height}.  In practice the beam would be blanked after each signal to reposition the nanostencil.

This method can also be applied to construct arrays of heavier donors implanted to a depth of 20~nm (see Table~\ref{tab:ions}).  The high-mass ions require higher kinetic energy to reach the same depth, but more e-h pairs are lost owing to recombination in the increasingly dense plume of ionization along the heavier ion tracks.

The limits of ion placement spatial precision for this method are given by the fundamental straggling inherent in the ion-solid interaction, the diameter of the aperture in the nanostencil and the sample stage positioning accuracy.  For 20 nm deep $^{31}$P implants at 14 keV these are 10 nm, sub-10 nm and sub-2 nm, respectively.  However, the higher mass donors have less straggling and hence potentially greater positioning precision (Table~\ref{tab:ions}).  For near-surface sub-20~nm deep implants it is also possible to make use of PF$_x$ molecular ions where the F bystander atoms create additional ionization that helps generating sufficient e-h pairs above the threshold for detection.  The likely diffusion of the F atoms away from the implant site during the post-implant annealing mitigates degradation of the donor quantum states. Theoretical studies suggest this method could be developed to have sufficient precision to fabricate viable devices \cite{vandonkelaar2010}. 

\begin{table}[t!]
\vspace{12 pt}

\begin{tabular}{ccccc}
\hline
Ion & Energy & Ionization (keV) & Electron-hole & Straggle\\
 & (keV) & (keV) & pairs & (nm) \\
 \hline \\
$^{31}$P & 14 & 3.5 & 950 & 10 \\
$^{75}$As & 23 & 4.0 & 1100 & 7 \\
$^{123}$Sb & 26 & 3.2 & 870 & 6 \\
$^{209}$Bi & 33 & 2.8 & 760 & 5\\
\hline
\end{tabular}

\caption{\label{tab:ions} The number of electron-hole pairs induced from donor ions implanted 20 nm deep in silicon.  Ionization values are calculated from the parameterization of Funsten et al.~\cite{funsten2004}.  The straggle is the standard deviation in the lateral position from SRIM~\cite{ziegler2010srim}.}
\end{table}

\section{\label{sec:conclusions}Conclusions and outlook}

In this article, we have summarized the current state of research on single-spin and small-ensemble implanted-donor devices in silicon, and provided some indications of near- and long-term directions for the field. 

On the topic of quantum information processing, donor spin qubits have proven to be highly coherent and high-fidelity systems, with record performance metrics among solid-state systems (Section~\ref{sec:benchmarks}). Several directions will be pursued to scale up this system, using exchange (Section~\ref{sec:exchange}) or electric dipole-mediated (Section~\ref{sec:flipflop}) 2-qubit logic gates, and spacing them out to accommodate interconnects and readout devices fabricated on length scales compatible with standard industry processes. Our current focus is on the flip-flop qubit proposal, which enables a 200~nm spacing between the qubits, consistent with the size and pitch of standard gates in modern MOS devices, while allowing the placement of 25 million physical qubits in a square millimeter. This is certainly not the only way forward: several alternative pathways exist for connecting donor qubits at various length scales (Section~\ref{sec:scaling}).

The main challenge for this endeavor is the ability to routinely fabricate devices with many individually addressable dopants, interfaced with classical control electronics. For the ion-implanted fabrication pathway described in this article, the challenges are not dissimilar from those facing the lithographically-defined semiconductor quantum dots. Interconnects and gate density issues can be studied, and hopefully resolved, in a holistic manner across both platforms \cite{vandersypen2017}. Integrating counted single-ion implantation (Section~\ref{sec:implantation}) within a CMOS process flow will be of paramount importance; existing proof-of-principle demonstrations \cite{singh2016} provide a tangible source of optimism.

Unlike artificially engineered systems such as quantum dots or superconducting qubits, donors are natural and identical atoms that can be meaningfully addressed in ensemble experiments. A small ensemble of donors can be used as a long-lived spin memory in a quantum Turing machine (Section~\ref{sec:ensembles}), providing a unique opportunity to develop hybrid systems for quantum information processing.

The atomic nature of the donors makes them appealing test beds for fundamental studies across diverse fields, such as quadurupole interactions (Section~\ref{sec:quadrupole}) and the striking discovery of nuclear electric resonance (Section~\ref{sec:NER}), to the search for a physical embodiment of a prototypical quantum chaotic system as the periodically-driven top (Section~\ref{sec:chaos}). In a more applied direction, donor spins can act as exquisite sensors for magnetic fields or mechanical strain (Section~\ref{sec:sensing}), where their atomic size and compatibility with semiconductor devices promises interesting applications.

The common theme of these research projects is the pursuit of useful (or simply interesting) quantum effects using well-defined atomic-size systems, integrated within the most important physical platform in the modern technological era: silicon MOS devices. Therefore, the potential for scaling and manufacturing ion-implanted donors in silicon underpins their permanence on the centre stage of quantum technologies for the foreseeable future.

\medskip
\textbf{Acknowledgements} \par

$^{\ast}$Email address: a.morello@unsw.edu.au

A.M. and D.N.J. acknowledge funding from the Australian Research Council Centre of Excellence for Quantum Computation and Communication Technology (project number CE170100012) and the US Army Research Office (grant number W911NF-17-1-0200). A.M. and J.J.P. acknowledge funding from an Australian Research Council Discovery Project (DP180100969). A.M. was supported by the Australian Department of Industry, Innovation and Science
(Grant no. AUSMURI00002). J.J.P. acknowledges an Australian Research Council Early Career Researcher Award (DE190101397). P.B. acknowledges support of the European Research Council under the European Community’s Seventh Framework Programme (FP7/2007-2013) through grant agreement No. 615767 (CIRQUSS), of the Agence Nationale de la Recherche under the Chaire Industrielle NASNIQ (grant number ANR-17-CHIN-0001), and of Region Ile-de-France Domaine d'Interet Majeur SIRTEQ under grant REIMIC. The views and conclusions contained in this document are those of the authors and should not be interpreted as representing the official policies, either expressed or implied, of the ARO or the US Government. The US Government is authorized to reproduce and distribute reprints for government purposes notwithstanding any copyright notation herein.

We acknowledge contributions to from W.D. Vine (image in Fig.~\ref{fig:ff_cavity}(b)), A.M. Jakob (data in Figure~\ref{fig:stencil} and \ref{fig:pulse_height}), S.E. Robson (Figure~\ref{fig:stencil} inset). 


\end{document}